\documentclass[useAMS,usenatbib,onecolumn]{mn2e}

\usepackage{amsmath,amssymb}
\usepackage{epsfig}
\usepackage{natbib}

\begin{document}

\title[Absorption variability in compact radio sources]{Absorption variability as a probe of the multiphase interstellar media surrounding active galaxies}

\author[Jean-Pierre Macquart and Steven Tingay]{Jean-Pierre Macquart$^{1,2}$\thanks{E-mail:
J.Macquart@curtin.edu.au} and Steven Tingay$^{1,2}$ \\
$^{1}$ICRAR/Curtin University, Curtin Institute of Radio Astronomy, Perth WA 6845, Australia\\
$^{2}$ARC Centre of Excellence for All-Sky Astrophysics (CAASTRO)}


\pagerange{\pageref{firstpage}--\pageref{lastpage}} \pubyear{2016}

\maketitle

\label{firstpage}

\begin{abstract}
We examine a model for the variable free-free and neutral hydrogen absorption inferred towards the cores of some compact radio galaxies in which a spatially fluctuating medium drifts in front of the source.  We relate the absorption-induced intensity fluctuations to the statistics of the underlying opacity fluctuations.  We investigate models in which the absorbing medium consists of either discrete clouds or a power-law spectrum of opacity fluctuations. We examine the variability characteristics of a medium comprised of Gaussian-shaped clouds in which the neutral and ionized matter are co-located, and in which the clouds comprise spherical constant-density neutral cores enveloped by ionized sheaths.  
The cross-power spectrum indicates the spatial relationship between neutral and ionized matter, and distinguishes the two models, with power in the Gaussian model declining as a featureless power-law, but that in the ionized sheath model oscillating between positive and negative values.  We show how comparison of the HI and free-free power spectra reveals information on the ionization and neutral fractions of the medium.  The background source acts as a low-pass filter of the underlying opacity power spectrum, which limits temporal fluctuations to frequencies $\omega \la \dot{\theta}_v/\theta_{\rm src}$, where $\dot{\theta}_v$ is the angular drift speed of the matter in front of the source, and it quenches the observability of opacity structures on scales smaller than the source size $\theta_{\rm src}$.  For drift speeds of $\sim 10^3\,$km\,s$^{-1}$ and source brightness temperatures $\sim 10^{12}\,$K, this limitation confines temporal opacity fluctuations to timescales of order several months to decades.
\end{abstract}


\section{Introduction}
There is an accumulating body of evidence to suggest that the lines of sight to the compact central regions of some active galaxies exhibit significantly variable absorption due to changes in the column of both neutral and ionized hydrogen.
Modelling of the broadband spectra of PKS\,1718-649 over the range 0.15-10\,GHz has revealed evidence for changes in its free-free opacity on timescales of months to years \citep{tin15,tin03}.  X-ray observations of the centres of the regions associated with AGN activity exhibit large ($>20$\%) changes in $N_H$ on timescales between days and weeks (\citealt{ris02}, see also \citealt{siem16}).   There is also some evidence for variations in the HI optical depths of some radio sources \citep{kan01,wol82}.

The interpretation of each of these individual absorption effects has hitherto been treated in isolation.  A common theme of existing models involves the passage of clouds across the line of sight to a source which is presumed to be relatively compact in angular extent.  \citet{bic97} considered a model for the spectrum of a source obscured by free-free absorption due to a population of clouds whose optical depths follow a power-law distribution, although they did not explicitly compute the statistics of the  temporal variations associated with their model.  \citet{ris02} present cloud-based models to account for the variable X-ray photoelectric absorption observed in several sources.  As the evidence for HI spectral variability is presently tenuous, there has been little systematic attempt to model it or to deduce its implications for AGN environments, however \citet{bri83} and \citet{mac05} suggest various cloud-based models as explanations.


The purpose of this paper is to present a unified framework for the interpretation of absorption variability towards the cores of radio AGN. The ultimate aim of such an approach is to deduce the composition of the multi-phase medium in the vicinity of active nuclei. This presents a means to ascertain what relation exists, if any, between the absorbing structures responsible for the HI, free-free and, ultimately, X-ray absorption.  At first glance the basis for the approach may appear unsound because the ionization fraction of the medium might be expected to change rapidly with distance from the nucleus, so that the structures which dominate each type of absorption are physically unconnected.  The rapid variability timescale of X-ray variations suggests that the absorption occurs very close to the central engine, whereas the month- to year-timescale HI and free-free variations are indicative of larger structures.  On the other hand, it is not clear-cut that the medium surrounding the central engine should be cleanly stratified into layers: the turbulent energy that exists in this environment should stir up the medium to prevent a clear delineation of the various phases, so that all phases are spatially co-existent.  

A rigorous framework provides the basis on which to interpret the opacity variations to make firm deductions about the veracity of either scenario.  In this paper we restrict our analysis to opacity variations at radio wavelengths for two reasons: (i) radio-based diagnostics themselves are sufficient to identify the relation between the distribution of neutral and ionized hydrogen, and (ii) there are likely to be strong differences in the structure and location of the X-ray and radio-emitting background sources which would hinder quantitative comparison of the absorption diagnostics.  Nevertheless, we remark that, although X-ray observations of AGN have been undertaken with day to week cadences, radio observations of spectral variability due to free-free and HI absorption processes have not been made with such high cadences.  This paper explores the motivation to investigate this observational regime.

The paper is organised as follows.  In Section \S \ref{sec:TheModel} we present a general model for absorption by an ensemble of clouds.  In Section \S \ref{sec:CloudsSection} we link the characteristics of the clouds with their ensemble-average variability properties.  We use this model to explore the significance of temporal variability between the various absorption-based radio-wavelength diagnostics.  In Section \S \ref{sec:Discussion} we demonstrate how the model may be used to glean information from existing variability data.  Our conclusions are presented in \S \ref{sec:Conclusions}.

\section{Temporal spectra of absorption fluctuations} \label{sec:TheModel}
\begin{figure*}
\epsfig{file=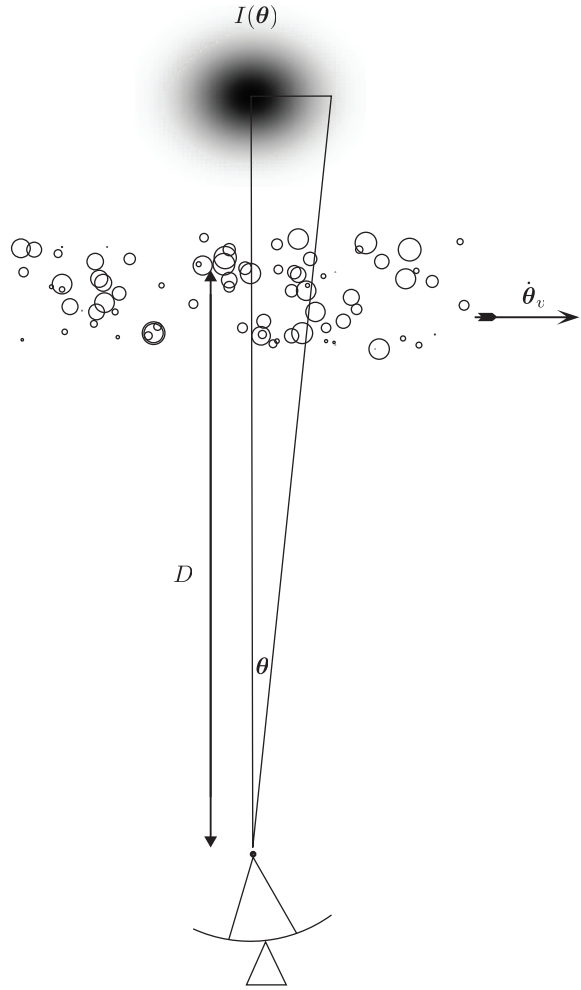,scale=0.5}
\caption{A depiction of the geometry of the absorption model.  The absorbing material is located on a thin plane at an angular diameter distance $D$ from the observer, and drifts across the line of sight at an angular velocity $\dot{\btheta}_v$.  The absorbing structure is view against a bright compact source with brightness distribution $I(\btheta)$.} \label{fig:Geometry}
\end{figure*}

In this section we discuss the power spectrum of temporal intensity fluctuations due to the passage of a distribution of absorbing material across the line of sight to a background source, as depicted in Figure \ref{fig:Geometry}.  Consider an absorbing medium at an angular diameter distance $D$ with an angular distribution of opacity $\tau(\btheta; \nu)$ across the sky that drifts at some angular speed $\dot{\btheta}_v$ across the line of sight to a source whose brightness distribution is $I(\btheta;\nu)$.  We approximate the opacity fluctuations to be ``frozen-in'' to a moving screen, which means that we take the motions of absorbing clouds relative to one another to be small compared to the speed at which the overall pattern traverses the line of sight.  The opacity distribution at a time $t$ may then be written in the form $\tau(\btheta - \dot{\btheta}_v t)$, and the time-dependent flux density observed at a frequency $\nu$ is
\begin{eqnarray}
S_\nu(t;\nu) = \int  I(\btheta;\nu) \exp \left[ -\tau(\btheta - \dot{\btheta}_v t ; \nu) \right] d^2 \btheta, \label{eq:Snu}
\end{eqnarray}
where we make the (excellent) assumption that the brightness temperature of the background source, $T_B$ far exceeds the spin temperature of the absorbing gas, $T_S$ (i.e. we assume that $T_S \ll T_B$), so that emission from the gas is negligible compared to the absorption.  This condition is easily satisfied for most species since the brightness temperature of the background source is expected to be within a few orders of magnitude of the Inverse Compton limit ($T_B \sim 10^{12}\,$K).

The foregoing formalism also admits in a restricted sense the treatment of opacity fluctuations associated with any relativistic synchrotron-emitting gas that may be present in the foreground of the radio source.  Specifically, we may examine the distribution of relativistic gas in the vicinity of the HI and free-free absorbing media using this formalism if the brightness temperature of the synchrotron-emitting gas is substantially less than that of the background source.  In this instance we may treat the fluctuations in the synchrotron optical depth in a similar manner to the free-free opacity variations in our formalism below.   


In general the exact angular distribution of $\tau$ is unknown {\it a priori}, but it is possible to describe the temporal flux density variations in terms of the corresponding power spectrum, or cross-power spectrum when comparing absorption at two different frequencies, $\nu$ and $\nu'$.  The power spectrum characterises the temporal variations associated with HI or free-free absorption alone.  The cross-power spectrum compares HI absorption intensity changes with intensity changes due to the free-free opacity.  The objective is to relate the temporal spectrum to the underlying statistical properties of the opacity distributions.  

In the following discussion it is useful to define the functions $C_\tau$, which measure the cross-correlation of the opacity fluctuations between frequencies $\nu$ and $\nu'$ on a scale $\btheta$.  These are related to the underlying power spectra and cross-power spectra of the opacity fluctuations, $\Phi_\tau ({\bf q};\nu,\nu') \equiv \langle \tilde{\tau}^*({\bf q}; \nu) \tilde{\tau}({\bf q};\nu') \rangle$, where $\tilde{\tau}({\bf q};\nu)$ represents the Fourier transform of $\tau(\btheta;\nu)$, as follows:
\begin{eqnarray}
C_\tau (\btheta,\nu,\nu') = \langle \tau(\btheta+\btheta',\nu) \tau(\btheta',\nu') \rangle = \frac{1}{(2 \pi)^2} \int d^2{\bf q} \, e^{i {\bf q} \cdot \btheta} \Phi_\tau ({\bf q};\nu,\nu'). \label{eq:autocovar}
\end{eqnarray}
The quantity ${\bf q}$ is the Fourier conjugate of ${\btheta}$ and is hence a dimensionless quantity. However, since the substitution ${\bf q} \rightarrow {\bf q}/D$ in $\Phi_\tau({\bf q},\nu,\nu')$ yields the power spectrum of spatial opacity fluctuations, we simply henceforth refer to $\Phi_\tau$ as the spatial power spectrum to distinguish it from the spectrum of temporal intensity fluctuations in the discussion below.

We thus compute the ensemble average spectrum of the flux density fluctuations:
\begin{eqnarray}
\langle {\cal P}(\omega; \nu,\nu') \rangle = \int  I (\btheta;\nu) I(\btheta';\nu') d^2 \btheta d^2\btheta' \int_{-\infty}^\infty dt dt' e^{i \omega (t-t')} \left\langle \exp \left[ - \tau(\btheta - \dot{\btheta}_v t ; \nu)  - \tau(\btheta' - \dot{\btheta}_v t' ; \nu') \right] \right\rangle. \label{tauPower}
\end{eqnarray} 
The average over the opacity variations is computed using the result $\langle \exp (-x) \rangle = \exp \left( \langle x^2 \rangle/2 \right)$, for a gaussian random variable\footnote{The assumption of normality has a proven pedigree  in a number of closely-related fields (e.g. for the statistics of the column density of ionized plasma in our Galaxy's interstellar medium (ISM)) in successfully predicting various observables (e.g. the scatter-broadening of compact radio sources).  As the HI optical depth is also linear in the matter density, it seems reasonable to adopt this approximation here.  To the extent that fluctuations in the ionized density of our own ISM follow a normal distribution, variations in $\tau_{\rm ff}$, though quadratic in the local density, might similarly be expected to exhibit statistics closely approximating that of a normal distribution, even if only by virtue of the central limit theorem.  Moreover, even when the opacity is not normally distributed, the result for $\langle \exp(-\tau) \rangle$ used here is still correct to second order in $\tau$.  Significant differences are only evident for $\tau \gg 1$, a condition unlikely to be relevant here (see text below).}, $x$.  We further assume that the opacity fluctuations are wide-sense stationary -- that is, that differences involving $\tau$ measured at different angular positions are only a function of the angular separation between those two positions -- so that eq.(\ref{tauPower}) becomes
\begin{eqnarray}
\langle {\cal P}(\omega; \nu,\nu') \rangle &=& \int I (\btheta;\nu) I(\btheta';\nu') d^2 \btheta d^2\btheta' \int dt dt' e^{i \omega (t-t')} \nonumber \\
&\null& \,\,  \times \exp \left\{\frac{1}{2} \left[ C_\tau(0;\nu,\nu) + C_\tau(0,\nu',\nu') + C_\tau(\btheta-\btheta' - \dot{\btheta}_v (t-t') ,\nu,\nu') +   C_\tau(\btheta'-\btheta + \dot{\btheta}_v (t-t') ,\nu',\nu)   \right] \right\}, \label{tauPower2}
\end{eqnarray}

Although the opacity itself may exceed unity, eq.(\ref{tauPower}) shows that only the fluctuations in $\tau$, which are expected to be much less than one, are of interest here.  This expectation is motivated by observations which show that the mean HI optical depth $\tau_{\rm HI}$ is typically much less than one \citep[Allison et al. in prep.]{kan01,wol82}.  Variations in the free-free opacity, $\tau_{\rm ff}$, may in principle be much larger than one at low sufficiently frequencies, so the approximation here is confined to frequencies in the regime $\tau_{\rm ff} \ll 1$.  This is satisfied in most cases of observational interest, since for $\tau_{\rm ff} \gg 1$, the sharp decline of the function $\exp(-\tau_{\rm ff})$ will render the source difficult to detect in this regime.  Granted this assumption, we are justified in linearising the exponential term in eq.(\ref{tauPower2}): 
\begin{eqnarray}
\langle {\cal P}(\omega; \nu,\nu') \rangle &=& \int I (\btheta) I(\btheta') d^2 \btheta d^2\btheta' \int dt dt'  e^{i \omega (t-t')} \int \frac{d^2{\bf q}}{(2 \pi)^2} \left[ 1 +\frac{1}{2} \Phi_\tau({\bf q};\nu,\nu) +\frac{1}{2} \Phi_\tau({\bf q},\nu',\nu') \right. \nonumber \\
&\null& \qquad \qquad \qquad \qquad  \left. 
+ \frac{1}{2} e^{i {\bf q} \cdot [\btheta-\btheta' - \dot{\btheta}_v (t-t')]} \Phi_\tau({\bf q},\nu,\nu')  + \frac{1}{2}  e^{-i {\bf q} \cdot [\btheta-\btheta' - \dot{\btheta}_v (t-t')]}  \Phi_\tau({\bf q} ,\nu',\nu)   \right]. \label{tauPower3}
\end{eqnarray}
Neglecting terms that contribute only at zero frequency and performing the Fourier transforms over time, the spectrum of temporal intensity fluctuations reduces to
\begin{eqnarray}
\langle {\cal P}(\omega; \nu,\nu') \rangle 
&=&    \int I (\btheta) I(\btheta') d^2 \btheta d^2\btheta'   \int \frac{d^2{\bf q}}{2} \, \delta(\dot{\btheta}_v \cdot {\bf q} + \omega) \left[ e^{i {\bf q} \cdot (\btheta - \btheta')} \Phi_\tau({\bf q};\nu,\nu') +  e^{-i {\bf q} \cdot (\btheta - \btheta')} \Phi_\tau^*({\bf q};\nu,\nu') \right] \nonumber \\
&=&  \frac{1}{2 \dot{\theta}_v}  \int I (\btheta;\nu) I(\btheta';\nu') \left[ F(\omega,\btheta,\btheta') + F^*(\omega,\btheta,\btheta') \right] d^2 \btheta d^2\btheta' \label{tauPowerFinal}, \\
&\null& \qquad \qquad \hbox{with } F(\omega,\btheta,\btheta') = e^{- i(\theta_\parallel-\theta_\parallel') \omega/\dot{\theta}_v} \int  e^{i q_\perp (\theta_\perp-\theta_\perp')} 
\Phi_\tau \left({\bf Q};\nu,\nu' \right) d q_\perp,
\end{eqnarray}
where we have decomposed ${\bf q}$, $\btheta$ and $\btheta'$  into components parallel and perpendicular to the drift vector $\dot{\btheta}_v$ and write ${\bf Q} = (-\omega/\dot{\theta}_v,q_\perp)$.

Equation (\ref{tauPowerFinal}) provides the means to relate the statistics of the intensity fluctuations associated with variable absorption directly back to the statistics of the opacity fluctuations themselves.  There is a particularly intuitive interpretation for the power spectrum of absorption fluctuations,
\begin{eqnarray}
\langle {\cal P}(\omega; \nu,\nu) \rangle &=&  
\frac{1}{\dot{\theta}_v} 
\left[ \int I (\btheta) 
e^{ i\theta_\parallel \omega/\dot{\theta}_v} d^2 \btheta \right] 
\left[ \int I(\btheta') 
e^{ -i\theta_\parallel' \omega/\dot{\theta}_v} d^2\btheta' \right] 
\int  e^{i q_\perp (\theta_\perp-\theta_\perp')} 
\Phi_\tau \left(\omega/\dot{\theta}_v,q_\perp;\nu,\nu' \right) d q_\perp  \nonumber \\
&=&
 \frac{1}{\dot{\theta}_v} \int \Phi_\tau \left(\omega/\dot{\theta}_v,q_\perp;\nu,\nu \right) d q_\perp
\left\vert \int I (\btheta;\nu) 
e^{ i\theta_\parallel \omega/\dot{\theta}_v + i q_\perp \theta_\perp} d^2 \btheta \right\vert^2. \label{tauPowerSimple}
\end{eqnarray}
First consider the fluctuations associated with a point-like source: the second integral reduces to a constant, and the power spectrum of intensity fluctuations is just the spectrum of opacity fluctuations, integrated over the component of the spectrum orthogonal to the drift direction. Thus we derive the intuitively obvious result that the temporal intensity fluctuations probes opacity structure only in the direction of the drift velocity.  

The effect of the integral over $I(\btheta)$ in eq.\,(\ref{tauPowerSimple}) is to act as a low-pass spectral filter.  This has two effects on the spectrum: it removes high frequency temporal fluctuations and it suppresses the overall amplitude of the variability.  A source of finite characteristic size $\theta_{\rm src}$ strongly attenuates frequencies $\omega \ga \dot{\theta}_v / \theta_{\rm src}$ in the temporal power spectrum.  One does not therefore expect opacity-induced intensity variations on timescales considerably less than the source crossing time $t_{\rm char} \sim \theta_{\rm src}/\dot{\theta}_v$.  Source size further attenuates the overall amplitude of the variability: a finite source only admits spatial frequencies $q_\perp \la \theta_{\rm src}^{-1}$, so that the integral over $q_\perp$ only samples the power contained in opacity fluctuations on angular scales larger than $\sim \theta_{\rm src}$.  

It is now straightforward to interpret the cross-power:
\begin{eqnarray}
\langle {\cal P}(\omega; \nu,\nu') \rangle   
&=&  \frac{1}{ \dot{\theta}_v} {\rm Re} \left\{ \int \Phi_\tau \left(-\omega/\dot{\theta}_v,q_\perp;\nu,\nu' \right) d q_\perp 
\left[ \int I (\btheta;\nu) 
e^{ -i\theta_\parallel \omega/\dot{\theta}_v + i q_\perp \theta_\perp} d^2 \btheta \right] \left[ \int I (\btheta';\nu') 
e^{ i\theta_\parallel' \omega/\dot{\theta}_v - i q_\perp \theta_\perp'} d^2 \btheta' \right] \label{tauCrossSimple}
\right\}.
\end{eqnarray}
In this case the cross power of the source structure measured at frequencies $\nu$ and $\nu'$ acts as a low pass filter to the cross-spectrum of opacity fluctuations\footnote{Of course, it is possible for HI and free-free absorption to occur at the same frequency, even though the foregoing theory treats their effects separately.  This approach is justified because the effect of the two is observationally separable: an observational analysis would separate HI absorption variations from free-free absorption absorption because the HI flux density variations are measured with respect to the nearby continuum spectrum (which may be subject to free-free absorption).}. 

In principle, equations (\ref{tauPowerSimple}) and (\ref{tauCrossSimple}) provide a complete formal means to investigate the nature of opacity variations in the absorbing medium.  It is possible to partially invert for the power spectrum and cross power of the opacity fluctuations and thus to determine the relationship between neutral hydrogen and ionized material in the medium surrounding a compact source.  To understand the extent to which one may actually invert the temporal variations, consider the intensity fluctuations in the simplest of cases, in which the background source is point-like with unit flux density. There are several measurable quantities: the power spectrum of HI intensity fluctuations, of free-free-induced intensity fluctuations and the cross power between these quantities.  The objective is to solve for the underlying statistics of the opacity induced by the neutral and ionized media, namely $\Phi_\tau({\bf q},\nu_{\rm HI},\nu_{\rm HI})$, $\Phi_\tau({\bf q},\nu_{\rm ff},\nu_{\rm ff})$ and $\Phi_\tau ({\bf q},\nu_{\rm HI},\nu_{\rm ff})$, where $\nu_{\rm ff}$ is a frequency at which free-free absorption is measurable.  However, complete inversion is not possible: the best we can measure is a quantity that integrates over one axis of the power spectrum of opacity fluctuations.  Written out explicitly, the resulting temporal power spectra are:
\begin{subequations}
\begin{eqnarray}
\langle {\cal P}(\omega; \nu_{\rm HI},\nu_{\rm HI}) \rangle &=& \frac{1}{ \dot{\theta}_v} \int \Phi_\tau(\omega/\dot{\theta}_v,q_\perp,\nu_{\rm HI},\nu_{\rm HI}) d q_\perp, \\
\langle {\cal P}(\omega; \nu_{\rm ff},\nu_{\rm ff}) \rangle &=& \frac{1}{ \dot{\theta}_v} \int \Phi_\tau(\omega/\dot{\theta}_v,q_\perp,\nu_{\rm ff},\nu_{\rm ff}) d q_\perp, \\
\langle {\cal P}(\omega; \nu_{\rm HI},\nu_{\rm ff}) \rangle &=& \frac{1}{ \dot{\theta}_v} {\rm Re} \int \Phi_\tau(\omega/\dot{\theta}_v,q_\perp,\nu_{\rm HI},\nu_{\rm ff}) d q_\perp.  
\end{eqnarray} \label{PtSrcRelations}
\end{subequations}
To the extent there is no cross-correlation in the underlying opacity fluctuations, no correlated variability will be observed.  We remark that one may equally investigate the foreground of synchrotron-absorbing relativistic gas as free-free absorption by substituting $\nu_{\rm ff}$ for $\nu_{\rm synch}$, subject to the caveats discussed below eq.(\ref{eq:Snu}). 

The underlying spectra of HI and free-free opacity fluctuations are unknown in general.  To gain further insight into the physics we introduce specific models for the structure of the medium in the following section.


\section{Absorption by an ensemble of clouds} \label{sec:CloudsSection}
The particular choice of a specific model, involving an ensemble of discrete clouds for the structure of the absorbing region, is adopted here because it connects in a theoretical sense to the work of \citet{bic97} in terms of the modelling of GPS/CSS radio spectra.  Some numerical simulations of such a scenario have also been explored, by \citet{sax05}.  The success of the theoretical and simulation approaches, against observational data, indicate that this choice has significant future promise that may be relevant for free-free, HI, and X-ray absorption, as similar suggestions have been made for X-ray absorption by \citet{ris02} and for HI absorption by \citet{bri83}.

Let us apply the foregoing results to the specific case of absorption by an ensemble of clouds.  We determine the power spectra $\Phi_\tau ({\bf q},\nu,\nu')$ for clouds comprised of an admixture of neutral and ionized material.  We parameterise the density associated with each cloud, given by a function $T({\btheta}-\Delta \btheta,z;\theta_{\rm c})$, in terms of its size $\theta_c$ and its angular offset, $\Delta \btheta$, from the line of sight at time $t=0$, where $z$ is the co-ordinate parallel to the line of sight.  

Since absorption due to neutral hydrogen is linearly proportional to the column density of material, we write the total opacity due to an ensemble of $N$ clouds as 
\begin{eqnarray}
\tau_{\rm HI}(\btheta) = \sum_j^N  A_j (\nu) \int_0^\infty T({\btheta}-\Delta \btheta_j,z;\theta_{\rm c,j}) dz \equiv \sum_j^N A_j(\nu) {\cal T}_{\rm HI} ({\btheta}-\Delta \btheta_j;\theta_{\rm c,j}),
\end{eqnarray}
where the column profile of each cloud, ${\cal T}_{\rm HI} ({\btheta}-\Delta \btheta_j;\theta_{\rm c,j})$, is defined by the above equation to be line-of-sight ($z$-axis) integral of the density, and the constant $A_j(\nu)$, with units of length-squared, relates the column profile ${\cal T}_j$ to the optical depth.  For absorption due to neutral hydrogen, the constants $A_j$ are
\begin{eqnarray}
A_j (\nu) = X_j \frac{3 c^2}{32 \pi} \frac{A_{10}}{\nu_{10}} \frac{h}{k T_s} \phi(\nu) = 2.59 \times 10^{-18} \phi(\nu) \, X_j \, \left( \frac{T_s}{10^3\,{\rm K}}\right)^{-1} \hbox{cm}^2, 
\end{eqnarray}
where $X_j$ is the neutral fraction of the $j$th cloud, $\phi(\nu)$ is the frequency profile of the HI absorption line, $T_s$ is the gas spin temperature, $A_{10} = 2.85 \times 10^{-15}\,$s$^{-1}$ and $\nu_{10} = 1420.4\,$MHz.  It is useful to express the frequency profile in terms of the effective velocity width of the line $\Delta  v_{\rm eff} = \int \phi(v) dv$ so that one has $\phi(\nu) = (\nu_{10} \Delta  v_{\rm eff}/c)^{-1}$ and we write 
\begin{eqnarray}
A_j(\nu) = 5.47 \times 10^{-24} X_j \, \left( \frac{\Delta  v_{\rm eff}}{100\,{\rm km\,s}^{-1}} \right)^{-1} \left( \frac{T_s}{10^3\,{\rm K}}\right)^{-1}  \hbox{cm}^2, \qquad  \frac{|\nu-\nu_{10}|}{\nu_{10}} \la \frac{\Delta  v_{\rm eff}}{2 c}. \label{Ajdefn}
\end{eqnarray}

Free-free absorption is proportional to the square of the density, so we write the optical depth as 
\begin{eqnarray}
\tau_{\rm ff} (\btheta) = \sum_j^N B_j (\nu) \int_0^\infty T^2 ({\btheta}-\Delta \btheta_j,z;\theta_{\rm c,j}) dz \equiv \sum_j^N B_j (\nu)  {\cal T}_{\rm ff} ({\btheta}-\Delta \btheta_j;\theta_{\rm c,j}).
\end{eqnarray} 
The quantity ${\cal T}_{\rm ff}$ is identified as being proportional the emission measure of the source, $\int n_e^2 dl$.  With ${\cal T}_{\rm ff}$ expressed in the conventional units of pc\,cm$^{-6}$, the constant of proportionality between the opacity and the emission measure, $B_j$, is approximately given by (\citealt{lang84}; see also \citealt{bic97}):
\begin{eqnarray}
B_j(\nu) = 3.28 \times 10^{-7} \, Y_j  \, \left(\frac{T_e}{10^4\,{\rm K}} \right)^{-1.35} \left( \frac{\nu}{1\,{\rm GHz}} 
\right)^{-2.1},  \label{Bjdefn}
\end{eqnarray}
where $T_e$ is the gas temperature and $Y_j = \int n_e^2 dl / \int n^2 dl$ is the ratio the column of the squared-density of ionized material over that for the total material for each particular cloud.   

The positions and sizes of the clouds are unknown, but we can compute an average over these quantities to determine the average power and cross-power spectrum of the optical depth.  In the following subsections we explore various averages over these quantities.  

\subsection{Average over cloud positions}
The average power spectrum of opacity fluctuations is
\begin{subequations}
\begin{eqnarray}
\Phi_\tau ({\bf q},\nu_{\rm HI},\nu_{\rm HI}) &=& {\Bigg \langle} \sum_{j,k}^N e^{i {\bf q} \cdot (\Delta \btheta_j - \Delta \btheta_k)} 
 A_j A_k \tilde{\cal T}_{\rm HI} ({\bf q};\theta_{\rm c,j}) \tilde{\cal T}_{\rm HI}^* ({\bf q};\theta_{\rm c,k}) {\Bigg \rangle}, \\
\Phi_\tau ({\bf q},\nu,\nu') &=& {\Bigg \langle} \sum_{j,k}^N e^{i {\bf q} \cdot (\Delta \btheta_j - \Delta \btheta_k)}
  B_j B_k \tilde{\cal T}_{\rm ff} ({\bf q};\theta_{\rm c,j}) \tilde{\cal T}_{\rm ff}^* ({\bf q};\theta_{\rm c,k}) {\Bigg \rangle}, \\
\Phi_\tau ({\bf q},\nu_{\rm HI},\nu') &=& {\Bigg \langle} \sum_{j,k}^N e^{i {\bf q} \cdot (\Delta \btheta_j - \Delta \btheta_k)} 
A_j B_k \tilde{\cal T}_{\rm HI} ({\bf q};\theta_{\rm c,j}) \tilde{\cal T}_{\rm ff}^* ({\bf q};\theta_{\rm c,k})  {\Bigg \rangle}.
\end{eqnarray}
\end{subequations}

It is reasonable to expect that the cloud offsets are statistically independent of their sizes, $\theta_{\rm c,j}$, and their neutral and ionization fractions, embodied in the constants $A_j$ and $B_j$. We perform the average over cloud positions assuming that they are distributed randomly\footnote{This treatment follows the formalism introduced in \citet{mac05} in another context, which we recap here for completeness, but to which we refer the reader for additional details.} over a solid angle $\Omega$, with $p_2(\Delta \btheta_j,\Delta \btheta_k)/\Omega^2$ being the joint probability of finding the $j$th cloud within the region $(\Delta \btheta_j, \Delta \btheta_j+d\btheta)$ and the $k$th cloud within the region $(\Delta \btheta_k, \Delta \btheta_k+d\btheta)$. We assume that the distribution of the cloud positions varies at most weakly over this region (i.e.~the average cloud density changes at most slowly over the entire region), so that the joint probability can be expressed as a function of separation, $p_2(\Delta \btheta_j,\Delta \btheta_k)/\Omega^2 \equiv p_2(\Delta \btheta_j -\Delta \btheta_k)/\Omega^2$, and is independent of the average cloud position ${\boldsymbol \Theta} = (\btheta_j+\btheta_k)/2$. The quantity $p_2$ is thus identified as the two-point correlation function between cloud positions.
The average over cloud positions simplifies to a set of $N$ self-terms and $N(N-1)$ cross terms   
\begin{eqnarray}
\left\langle \sum_{j,k}^N e^{i {\bf q} \cdot (\Delta \btheta_j - \Delta \btheta_k)} \right\rangle &=& N + N (N-1) \left\langle  e^{i {\bf q} \cdot (\Delta \btheta_j - \Delta \btheta_k)} \right\rangle_{j \neq k} \nonumber \\
&=& N + \frac{N(N-1)}{\Omega^2} \int_{-\infty}^{\infty} d{\boldsymbol \Theta} \, d (\Delta \btheta_{j} - \Delta \btheta_{k}) \, p_2 (\Delta \btheta_{j} - \Delta \btheta_{k}) \, e^{i {\bf q} \cdot (\Delta \btheta_{j} - \Delta \btheta_{k})} \nonumber \\
&=& N + \frac{N(N-1)}{\Omega} \tilde{p_2} ({\bf q}),  \label{SpectrumPosnAvg}
\end{eqnarray}
where $\tilde{p_2}({\bf q})$ is the Fourier transform of $p_2(\Delta \btheta_{j} - \Delta \btheta_{k})$.  This average has a simple interpretation in terms of the cloud clustering statistics.   The two-point correlation function represents the effect of cloud clustering: if the cloud positions are distributed completely randomly, the two-point correlation function is identically zero and the $N(N-1)$ cross terms do not contribute.  However, if there is a tendency for clouds to cluster together, the two-point correlation function will fall to zero over a length scale comparable to the clustering scale, say $\theta_{\rm cluster}$, and the cross-terms will only contribute to the average over cloud positions on spatial frequencies $q \la \theta_{\rm cluster}^{-1}$.

The important result of this section is that, if clustering is unimportant, only the self-terms of each cloud contribute on average to the power spectrum of intensity variations.  

\subsection{Specific cloud models}

\subsubsection{Gaussian profile} 
\begin{figure}
\centerline{\epsfig{file=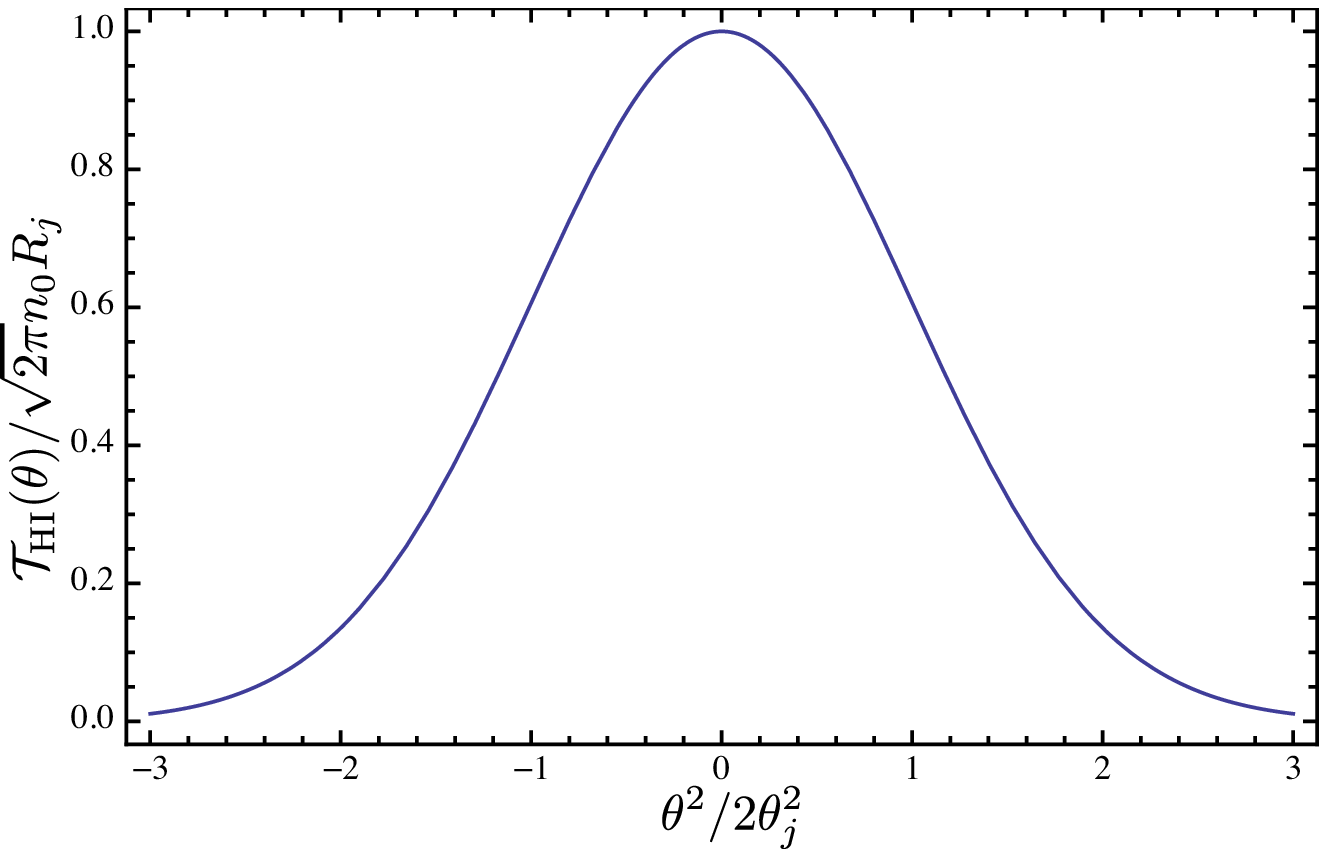,scale=0.5} \epsfig{file=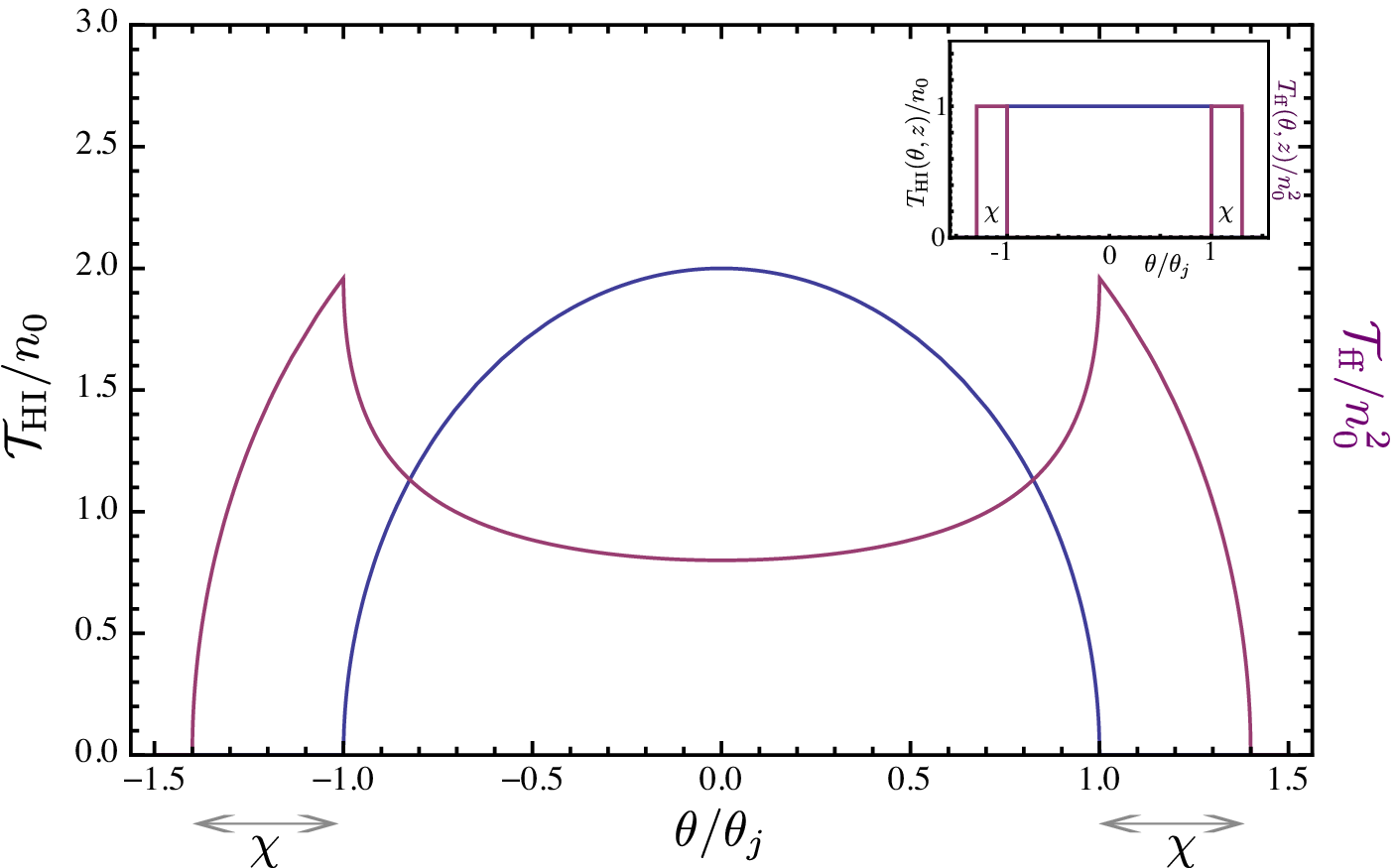,scale=0.5}}
\caption{The column profiles of the two models investigated here.  Left: the HI column density for the Gaussian cloud model (with the ionized component following the same functional form). Right: the column densities of the neutral and ionized components of the shell model, with (inset, top right) a plot of the corresponding angular dependence of $T_{\rm HI}(\theta,z)$ and $T_{\rm ff}(\theta,z)$ from which the column profiles are derived.} \label{fig:cloudProfiles}
\end{figure}

\begin{figure}
\epsfig{file=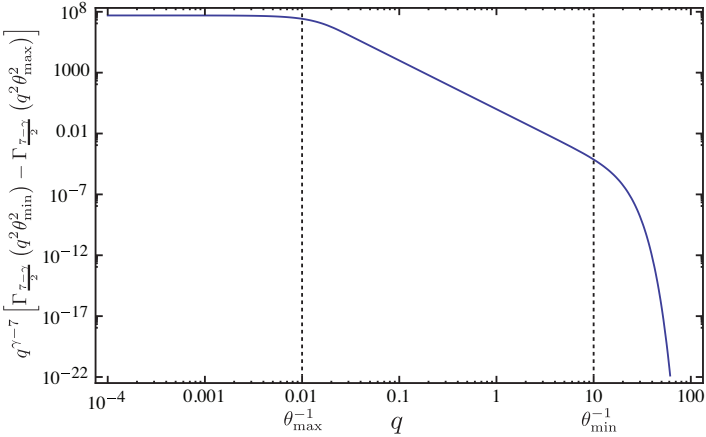,scale=0.7}
\caption{The generic behaviour of the spectra in equations \ref{GaussianSpectra}, here plotted with $\gamma=3$, $\theta_{\rm max} = 100$ and $\theta_{\rm min} = 0.1$.} \label{fig:gaussianGeneric}
\end{figure}

Let us neglect any potential effect of cloud clustering and investigate the power spectrum when the density profiles of the clouds follow a Gaussian form. This profile possesses the virtue of being highly analytically tractable while still capable of illustrating the important aspects of the underlying physics.  For clouds with with characteristic sizes $R_j= \theta_{c,j} D$, the density profile for clouds viewed along the $z$-axis is:
\begin{eqnarray}
T(\btheta,z;\theta_{c,j} ) = n_0 \exp \left[ - (D^2 \theta^2 +z^2)/2 R_j^2 \right],
\end{eqnarray}
where $D$ is the angular-diameter distance from the observer to the clouds.  The salient features of this model are illustrated in Fig.\,\ref{fig:cloudProfiles}.  The peak density of each cloud, $n_0$, is identical so that column depth variations are due only to the extent of each cloud and variations in the neutral and ionization fractions.  Neglect of cloud clustering means that only the $N$ self-terms contribute to the opacity, so the power spectrum of opacity fluctuations reduces to
\begin{eqnarray}
\Phi_\tau ({\bf q},\nu,\nu') 
&=& \pi^3 N  \begin{cases}
8  n_0^2 \left\langle A_j^2 R_j^2 \theta_{c,j}^4  \exp \left[ - \theta_{\rm c,j}^2 q^2 \right] \right\rangle, & \nu = \nu' = \nu_{\rm HI}, \\
 n_0^4 \left\langle B_j^2   R_j^2 \theta_{c,j}^4  \exp \left[ - \theta_{\rm c,j}^2 q^2/2 \right] \right\rangle, &  \nu = \nu' \neq \nu_{\rm HI}, \\
2^{3/2} n_0^3 \left\langle A_j B_j  R_j^2 \theta_{c,j}^4  \exp \left[ - 3 \theta_{\rm c,j}^2 q^2/4 \right] \right\rangle, & \nu =\nu_{\rm HI} \neq \nu', 
\end{cases}
\end{eqnarray}
where the power spectrum is a function of the dimensionless variable ${\bf q}$, the Fourier conjugate of $\btheta$.  If the neutral and ionization fractions of each cloud are independent of size, it is straightforward to perform the average over the cloud sizes.  For a power-law distribution of cloud sizes with index $-\gamma$, which ranges between minimum and maximum cloud sizes $\theta_{\rm min}$ and $\theta_{\rm max}$ respectively, the corresponding normalised size probability distribution is  
\begin{eqnarray}
p(\theta_{\rm c}) = K \theta_{\rm c}^{-\gamma}, \qquad K = \frac{1-\gamma}{\theta_{\rm max}^{1-\gamma} - \theta_{\rm min}^{1-\gamma}}.
\end{eqnarray}
The spectrum of HI and free-free absorbing material reduces to, for $\gamma < 7$,
\begin{subequations}
\begin{eqnarray}
\Phi_\tau ({\bf q},\nu_{\rm HI},\nu_{\rm HI})  &=& 4 N D^2 n_0^2 K \pi^3 q^{\gamma - 7}  
 \langle A_j^2 \rangle \left[ \Gamma_{\frac{7-\gamma}{2}} \left( q^2 \theta_{\rm min}^2 \right) - \Gamma_{\frac{7-\gamma}{2}} \left( q^2 \theta_{\rm max}^2 \right) \right], \\
 \Phi_\tau ({\bf q},\nu_{\rm ff},\nu_{\rm ff})  &=& 2^{\frac{5-\gamma}{2}} N D^2 n_0^4 K \pi^3 q^{\gamma - 7} 
 \langle B_j^2 \rangle \,  \left[ \Gamma_{\frac{7-\gamma}{2}} \left( \frac{q^2 \theta_{\rm min}^2 }{2} \right) - \Gamma_{\frac{7-\gamma}{2}} \left( \frac{q^2 \theta_{\rm max}^2 }{2} \right) \right] ,  \\
\Phi_\tau ({\bf q},\nu_{\rm HI},\nu_{\rm ff})  &=& 2^{15/12 - \gamma} 3^{\frac{\gamma-7}{2}} N D^2 n_0^3 K \pi^3 q^{\gamma - 7} 
 \langle A_j B_j \rangle \, \left[ \Gamma_{\frac{7-\gamma}{2}} \left( \frac{3 q^2 \theta_{\rm min}^2}{4} \right) - \Gamma_{\frac{7-\gamma}{2}} \left( \frac{3 q^2 \theta_{\rm max}^2}{4} \right) \right] . 
\end{eqnarray} \label{GaussianSpectra}
\end{subequations}
The generic nature of each spectrum, as illustrated in Figure \ref{fig:gaussianGeneric} is as follows: the spectrum is flat for $0<q<\theta_{\rm max}^{-1}$, follows a power-law decline with index $\gamma-7$, and then the spectrum declines sharply for $q>\theta_{\rm min}^{-1}$.  
The simplification enabled by neglect of clustering thus offers a straightforward means of extracting the parameters $A$ and $B$ which encapsulate the physics of the neutral and ionized medium fractions within each cloud, as we discuss further in \S \ref{sec:Discussion}.

\subsubsection{Neutral clouds with ionized shells} \label{subsec:IonizedSheath}

Simulations of jet environments, e.g. \citet{sax05}, can model the medium in the vicinity of an AGN as composed of clouds which are abraded by the AGN jet.  We are thus motivated  to consider the opacity fluctuations associated by an ensemble of clouds whose neutral hydrogen cores extend to some radius $R$ and whose ionized outer layer extends in a thin shell from a radius $R$ to a radius $R (1+\chi)$. 

\begin{figure}
\epsfig{file=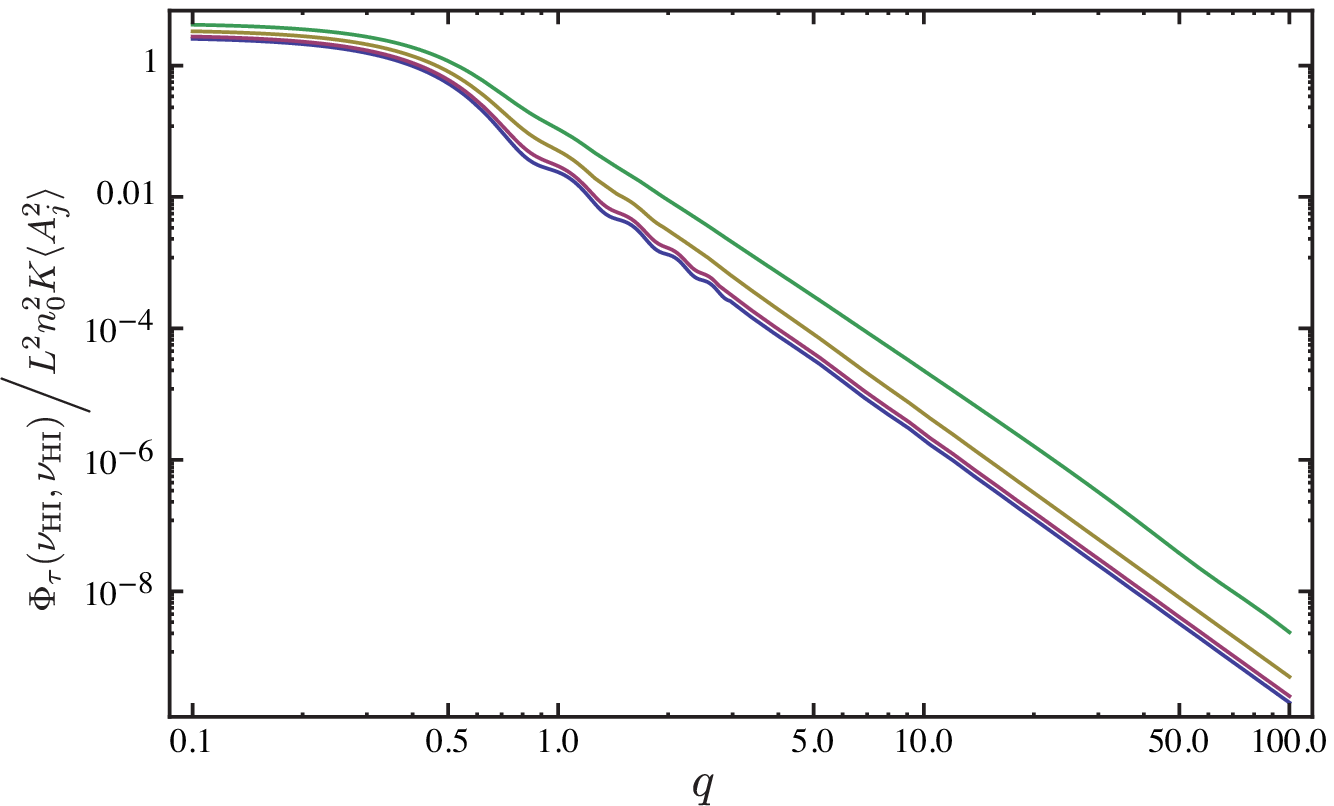,scale=0.7} \\
\epsfig{file=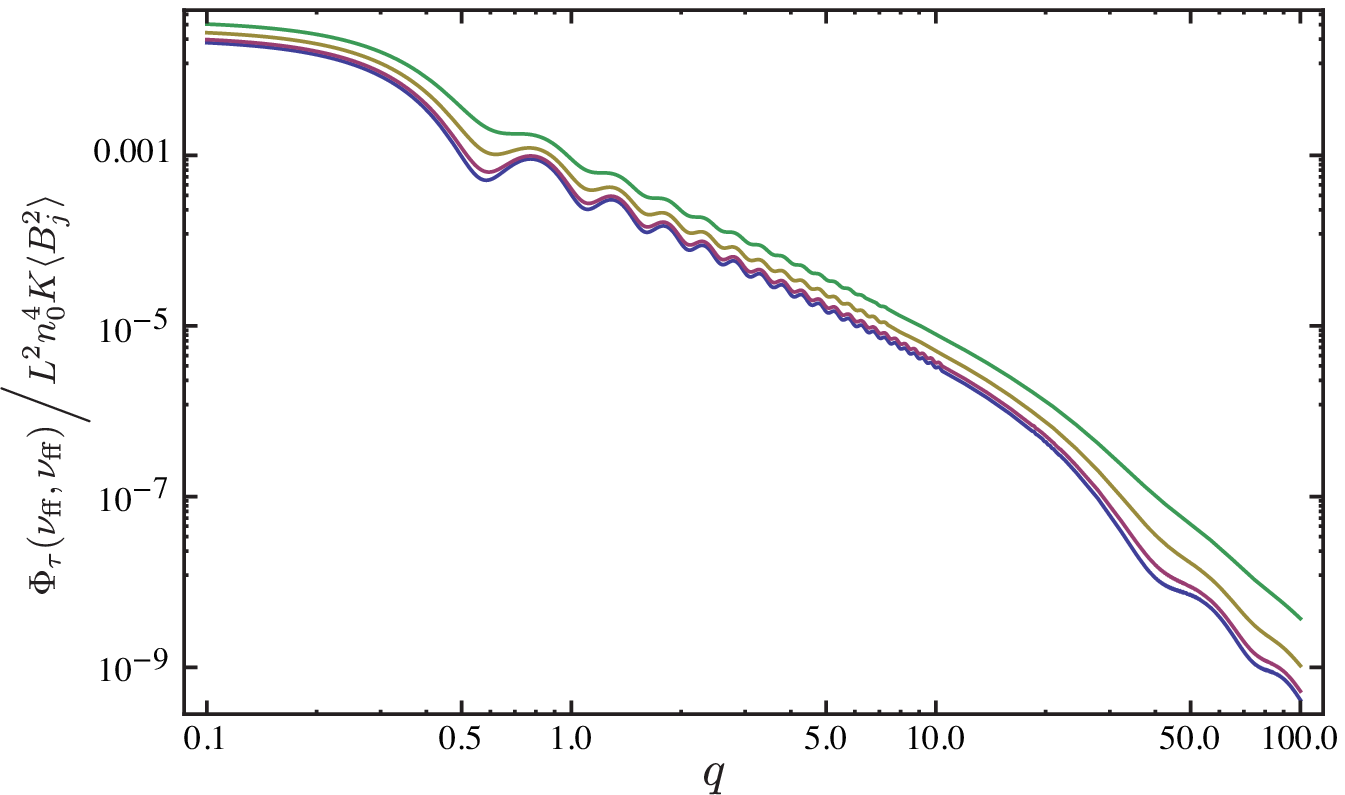,scale=0.7} \\
\epsfig{file=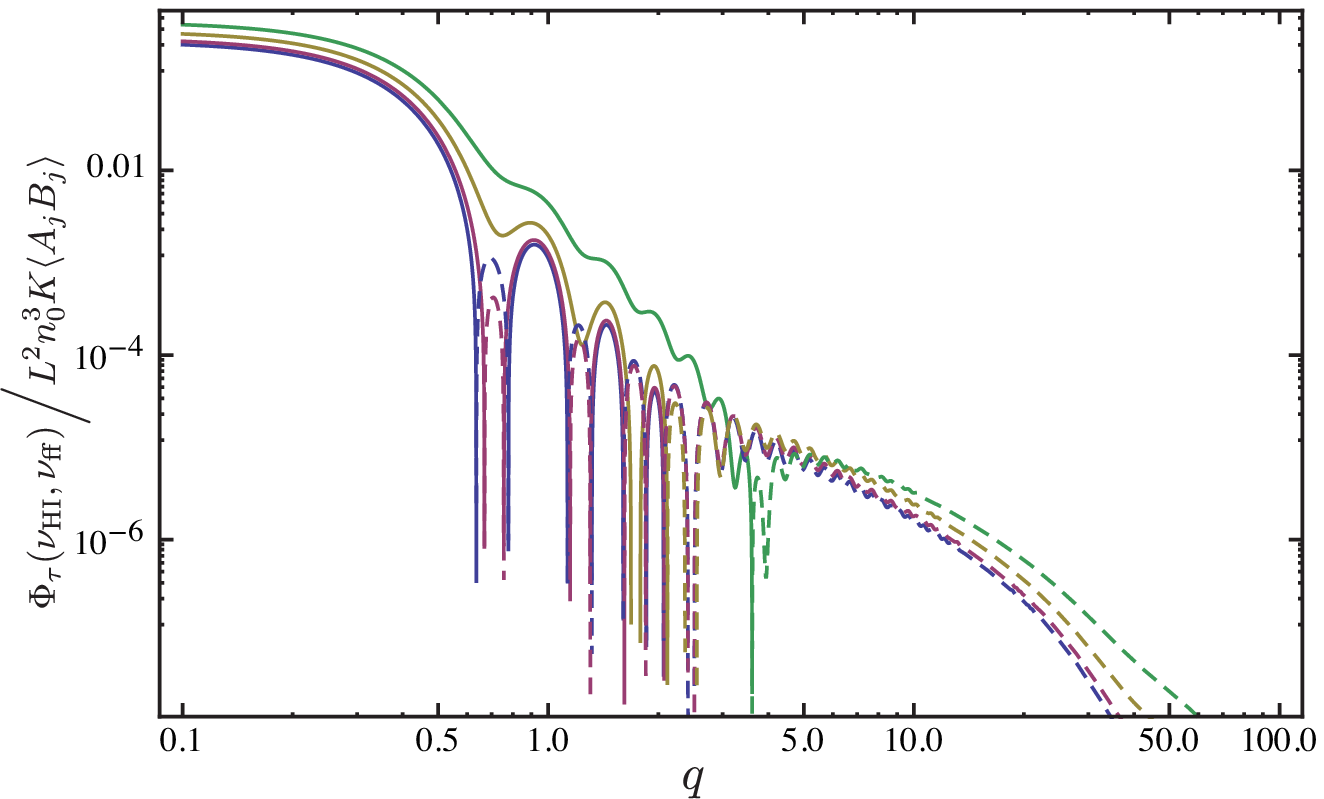,scale=0.7} 
\caption{The HI power spectrum, free-free power spectrum, and HI-free-free cross-power spectrum for the ionized sheath model.  The spectra are plotted here with parameters $\theta_{\rm min} = 0.01$, $\theta_{\rm max} =1.0 $ and $\chi=0.03$.  In the cross-power spectrum negative values are denoted by dotted lines.  The different curves denote the spectra for $\gamma=0.5,1.01,2$ and $3.01$  (in order of increasing amplitude).} \label{fig:SheathModel} 
\end{figure}

The HI and ionized profiles are therefore expressed in terms of the unit step function, $H(x)$, with density profiles $T_{\rm HI} (\btheta,z) = n_0 H [R^2 - (\theta^2 D^2 + z^2)]$ for the neutral matter, and with a density distribution of ionized material described by $T_i(\btheta,z) = n_0 \{H [R^2(1+\chi)^2 - (\theta^2 D^2 + z^2)] - H [R^2 - (\theta^2 D^2 + z^2)] \}$.  Integrating over the depth of the cloud, we obtain the HI and free-free column densities
\begin{subequations}
\begin{eqnarray}
{\cal T}_{\rm HI}(\btheta) &=& 2 n_0 \sqrt{R^2 - \theta^2 D^2} \, H[R^2 - \theta^2 D^2] \\
{\cal T}_{\rm ff}(\btheta) &=& 2 n_0^2 \left[ \sqrt{R^2 (1+\chi)^2 - \theta^2 D^2} \, H[R^2(1+\chi)^2 - \theta^2 D^2] -  \sqrt{R^2 - \theta^2 D^2} \, H[R^2 - \theta^2 D^2] \right]. 
\end{eqnarray}
\end{subequations}
These functions are plotted in Fig.\,\ref{fig:cloudProfiles}.  Their corresponding Fourier transforms, which depend only on the magnitude of ${\bf q}$ are, for a given cloud size $\theta_c$, 
\begin{subequations}
\begin{eqnarray}
\tilde{\cal T}_{\rm HI}({\bf q};\theta_c) &=& \frac{ D \theta_c \, n_0}{2 \pi^2 q^3} \left[ \frac{\sin (2 \pi q \theta_c)}{\theta_c} - 2\pi q \cos (2 \pi q \theta_c) \right],  \\
\tilde{\cal T}_{\rm ff}({\bf q};\theta_c) &=& \frac{D \theta_c \, n_0^2}{2 \pi^2 q^3} \left[\frac{ \sin (2 \pi q \theta_c (1+\chi)) -\sin (2 \pi q \theta_c) }{\theta_c} - 2\pi q \left[ (1+\chi) \cos (2 \pi q \theta_c(1+\chi)) - \cos (2 \pi q \theta_c) \right] \right]. 
\end{eqnarray}
\end{subequations}
These expressions may be integrated over a power law distribution of cloud radii to obtain the ensemble-average power- and cross-spectra of the opacity.  Although the model is simple, the analytic forms of the resulting power spectra are algebraically cumbersome, so we present them in Appendix \ref{app:Sheath} and describe the results in general terms here.  The general behaviour of the spectra is shown in Figure \ref{fig:SheathModel}, and is described as follows:  
\begin{itemize}
\item The HI and free-free power spectra are flat over the range $0<q \la 0.5 \, \theta_{\rm max}^{-1}$, where we recall that $\theta_{\rm max}$ is the maximum cloud size.
\item For $\gamma < 3$ the power spectrum declines as $q^{-4}$ for $q> \theta_{\rm max}^{-1}$.  For size distributions with $\gamma > 3$, the spectrum is dominated by $\theta_{\rm min}$, and it declines as $q^{-4}$ once $q > \theta_{\rm min}^{-1}$.  
\item The presence of the thin-shelled structure in the free-free angular distribution introduces ringing in the power spectrum with a characteristic ripple length $q \sim [\theta_{\rm max} (1+\chi)]^{-1}$.
\item The cross-power spectrum is approximately flat for $q \la 0.5 \,\theta_{\rm max}^{-1}$, but for $\gamma<3$ and $q \ga 0.5\, \theta_{\rm max}^{-1}$ the cross power oscillates between positive and negative values, and is bounded by an envelope whose amplitude scales as $q^{-4}$.  
\end{itemize}

\subsection{Temporal behaviour for power-law opacity spectra} \label{sec:TempSpec}
Having derived the two-dimensional power and cross-power spectra of the opacity-induced intensity fluctuations, we would like to relate these back to the observables, namely the spectra of temporal intensity fluctuations.  

We are motivated to consider the temporal power spectrum resulting from opacity fluctuations whose power spectrum is flat to some inner wavenumber, $q_{\rm min}$, corresponding to the inverse of the outer scale of the fluctuations, and which follows a power law for $q>q_{\rm min}$,
\begin{eqnarray}
\Phi_\tau({\bf q}) = \begin{cases} 
Q_0 \, q_{\rm min}^{-\beta},  & q \leq q_{\rm min}, \\
Q_0 \, q^{-\beta}, & q>q_{\rm min}. \\
\end{cases} \label{GenericPowerLaw}
\end{eqnarray}
This form encapsulates the behaviour of the HI and free-free power spectra derived in the two foregoing subsections.  It also captures the form of the cross-power spectrum for clouds with Gaussian profiles.  However, the form is more generally motivated by observations of the distribution of HI and ionized ISM in our own Galaxy.  Observations of the fine-scale structure in the neutral hydrogen content of our both Galaxy and neighbouring galaxies show that its distribution follows a power law with index $\sim-3$ \citep{cro83,sta99}.  Studies of the ionized component of our Galaxy's ISM indicate that the ionized ISM follows a power law with an index close to $\approx -11/3$ \citep{arm95}.  Our results below apply in general to any such medium whose power spectrum follows a power law of the type embodied in eq.(\ref{GenericPowerLaw}).

The parameters $Q_0$ and $q_{\rm min}$ relate back to the total variance in opacity fluctuations.  The cross-covariance in $\tau$ on the transverse angular scale ${\btheta}$ is related to the power spectrum by eq.(\ref{eq:autocovar}), from which the variance in the opacity fluctuations associated with a power spectrum of the form in eq.(\ref{GenericPowerLaw}) is obtained:
\begin{eqnarray}
\langle \delta \tau^2 \rangle = \frac{Q_0}{4 \pi} \frac{\beta}{\beta-2} q_{\rm min}^{2-\beta}.
\end{eqnarray}
 
As discussed in \S \ref{sec:TheModel}, the derivation of the spectrum of temporal intensity fluctuations involves integration of the product of the power (or cross-) spectrum with the source brightness power (cross-) spectrum over the co-ordinate orthogonal to the drift velocity, $q_\perp$ (see eqns.\,(\ref{tauPowerSimple}) \& (\ref{tauCrossSimple})).  Although derivation of the temporal power spectrum is complicated by its dependence on source size, which is in principle not known in detail, it is a possible to incorporate the effect of source size by approximating the brightness profile with a Gaussian form with characteristic size $\theta_{\rm src}(\nu)$, which we write explicitly as a function of observing frequency $\nu$.

For a source with a brightness profile $I(\btheta;\nu) = I_0(\nu) \exp[ -\theta^2/\theta_{\rm src}^2(\nu) ]$, the Fourier-transformed brightness is of the form $\tilde{I}({\bf q};\nu) = {\cal I}(\nu) \exp[-q^2 \theta_{\rm src}^2(\nu)/4]$, with ${\cal I}(\nu) = I_0(\nu) \, \pi \theta_{\rm src}^2(\nu)$.  Thus the spatial power spectrum cuts off at $q_{\rm src}^2 =2/\theta_{\rm src}(\nu)^2$, and the cross-spectrum measured between frequencies $\nu$ and $\nu'$ cuts off at  $q_{\rm src}^2 = 4/[\theta_{\rm src}^2(\nu) + \theta_{\rm src}^2(\nu')]$.

\begin{figure}
\epsfig{file=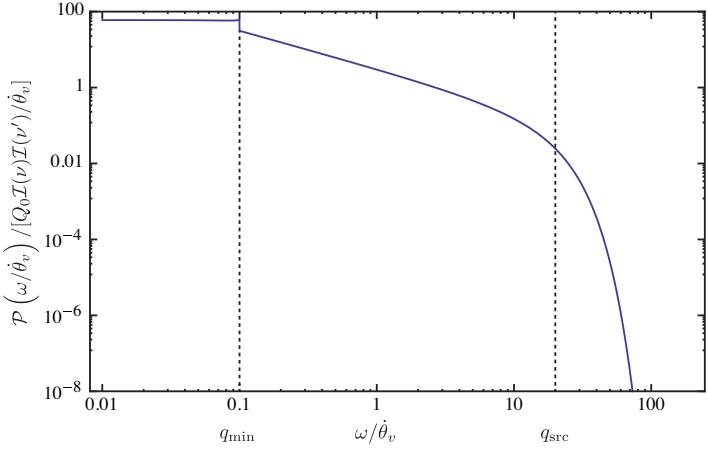,scale=0.7}
\caption{An illustration of the temporal power spectrum associated with an opacity spatial power spectrum whose index is $\beta=2$, plotted here for an opacity power spectrum whose low-spatial frequency cutoff is $q_{\rm min}=0.1$ and for a source size which cuts the spatial power spectrum off at $q_{\rm src}=20$.} \label{fig:TempSpec}
\end{figure}

The detailed calculation of the temporal power spectrum is presented in Appendix \ref{app:PowSpec}, and we summarise the results here.  The generic shape of the spectrum is shown in Figure \ref{fig:TempSpec}.  The temporal power spectrum is approximately flat for angular frequencies $0 < \omega/\dot{\theta}_v < q_{\rm min}$ and then declines as $\omega^{1-\beta}$ until the source size cuts the spectrum off near $\omega /\dot{\theta}_v \approx q_{\rm src}$.   The power-law portion of the spectrum is not evident if the source is sufficiently large, $q_{\rm src} < q_{\rm min}$ (i.e.\,\,if the angular size of the source exceeds the the outer scale of the fluctuations in optical depth).

\subsubsection{Temporal cross-spectrum for clouds with ionized sheaths}

The behaviour of the cross-spectrum for the ionized sheath model for an arbitrary distribution of cloud sizes is algebraically involved.  However, all the salient features of the correlation of the HI- and free-free absorption-induced temporal variations can be gleaned by restricting the scope of the examination to a specific distribution of cloud sizes.  We examine the simple case in which the cloud sizes follow the probability distribution $p(\theta_c) = K \theta_c^{-2}$ and the background source is point-like (i.e. $q_{\rm src} \rightarrow \infty$), so that the only free parameters are $\theta_{\rm min}$, $\theta_{\rm max}$ and the cloud edge thickness, $\chi$. The cross-correlation of opacity-induced intensity fluctuations has the solution:
\begin{eqnarray}
\Phi_\tau ({\bf q};\nu_{\rm HI},\nu_{\rm ff}) &=& \frac{D^2 n_0^3 N K \langle A_j B_j \rangle}{4 \pi^4 q^6} {\Bigg [} 
\sin [2 \pi q \theta_c (1+\chi)] \frac{2 \pi q \theta_c (1+\chi)^2 \cos (2 \pi q \theta_c) - \chi (2+\chi) \sin (2 \pi q \theta_c)}{\theta_c \chi (2+\chi)} \nonumber \\  &\null& \qquad
- \frac{2 \pi q (1+\chi) \cos [2 \pi q \theta_c (1+\chi)] \sin (2 \pi q \theta_c) }{\chi (2 + \chi)}  
- \frac{4 \pi^2 q^2 - 1 + \cos(4 \pi q \theta_c) + \pi q \theta \sin (4 \pi q \theta_c)}{2 \theta}
 {\Bigg ]}_{\theta_c = \theta_{\rm min}}^{\theta_c = \theta_{\rm max}}.
\end{eqnarray}
The behaviour of this function is dominated by the value of $\theta_{\rm max}$, and is insensitive to the value of $\theta_{\rm min}$ for $\theta_{\rm max} \gg \theta_{\rm min}$.  

\begin{figure}
\epsfig{file=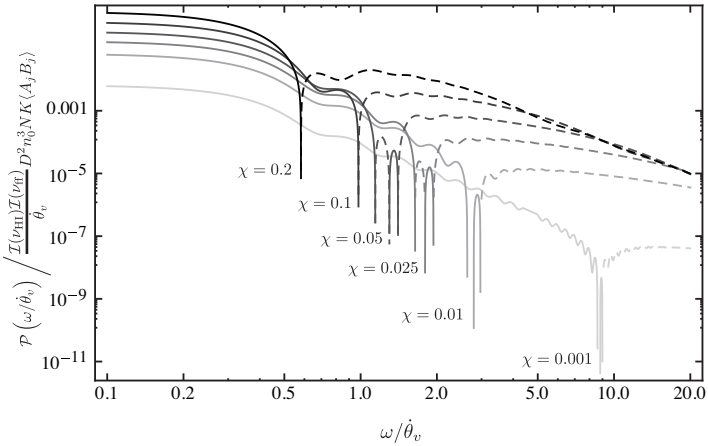,scale=0.7}
\caption{The temporal cross-power spectrum of variations between HI and free-free absorption in the ionized sheath model for a distribution of cloud sizes proportional to $\theta^{-2}$, with a maximum cloud size of $\theta_{\rm max}=1.0$ and a minimum size of 
$\theta_{\rm min}=0.01$, for various fractional sheath thicknesses $\chi$.} \label{fig:TempCrossCorr}
\end{figure}

The resulting temporal cross-power spectrum, obtained using eq.(\ref{PtSrcRelations}), is plotted for various values of $\chi$ in Figure \ref{fig:TempCrossCorr}.  It is seen that the spectrum is flat for $q \la (2 \pi \theta_{\rm max})^{-1}$, and the spatial cross-power spectrum possesses ripples that oscillate on a scale $q \approx \theta_{\rm max}^{-1}$, which are evident over the general decline in amplitude as $q \rightarrow \infty$: the amplitude of the spatial cross-spectrum, $\Phi_\tau$, declines as $q^{-4}$, and the amplitude of the temporal spectrum therefore declines as $q^{-3}$.   At sufficiently high values of $q$ the spatial cross-power spectrum becomes negative; in other words, there is a negative correlation between the HI and free-free structure on very fine scales.  This is to be expected given that the neutral and ionized media are not co-located within each cloud.  The lowest spatial wavenumber at which this occurs is obtained by expanding $\Phi_\tau ({\bf q};\nu_{\rm HI},\nu_{\rm ff})$ to second order in $\chi$, and we find that the zero-crossing points of the cross-spectrum are well-approximated by the solution to the equation
\begin{eqnarray}
(\theta_{\rm max} q)^2 = \frac{3}{8 \pi^2 \chi} \left[ 2 (\chi +2) + (2-\chi)  \cos 4 \pi q \theta_{\rm max}   - 2 \pi \chi q \theta_{\rm max}  \sin 4 \pi q \theta_{\rm max}  - \frac{\sin 4 \pi q \theta_{\rm max} }{2 \pi q \theta_{\rm max} } \left( 3 - \frac{\chi}{4} \right)  \right] \approx \frac{3}{4 \pi^2 \chi} \left( 2 + \cos 4 \pi q \theta_{\rm max} \right)  , 
\end{eqnarray}
where in the second approximation we have taken into account the fact that the solution for $\chi \ll 1$ occurs at $q \theta_{\rm max} >1$ to neglect the terms proportional to $1/(q \theta_{\rm max})$ inside the square brackets.  The important point here is that the zero crossing location is inversely proportional to the square-root of the cloud fractional thickness (i.e. $q \propto \chi^{-1/2}$) and inversely proportional to the maximum cloud size, $\theta_{\rm max}$.  The location of this first zero in the temporal cross-power spectrum is approximately equal to the value of $\omega/\dot{\theta}_v$ at which the spatial cross-spectrum flips from positive to negative values.

\section{Discussion} \label{sec:Discussion}
In principle measurements of the temporal spectra of opacity variations due to HI and free-free absorption in front of compact radio sources offer a means of discerning the fine-scale spatial structure of the media in external galaxies, and the cross-power spectrum offers a means of discerning the relationship between the spatial distribution of their neutral and ionized phases.  

However, two factors hamper our ability to determine the details of the spatial distribution from temporal opacity variations: the first is that the temporal variations integrate over the spectrum of spatial fluctuations in the direction orthogonal to the drift velocity.  This partially smooths over any features in the spectrum that are indicative of the detailed distribution of absorbing structure: the amplitude of features in the temporal spectrum is reduced relative to that in the spatial spectrum.  

The second effect is that of the finite size of the background source, which cuts off the temporal power spectrum at an angular frequency $\omega = \dot{\theta}_v /\theta_{\rm src}$.  The background source effectively acts as a low-pass filter of spatial information. This latter effect poses a serious limitation to measurements of the interrelation between neutral and ionized matter if the differences in their distribution occur only on angular scales finer than $\theta_{\rm src}$.  Our examination of the results of two contrasting cloud models introduced above illustrates this point.

We have considered several models in which absorbing clouds are distributed randomly without clustering and whose sizes follow a power law distribution with index $-\gamma$.  In the Gaussian cloud model the density of ionized and neutral matter both decline as a Gaussian from the cloud centre (i.e. the two phases are inter-mixed on tiny scales), and the spectra of temporal flux density variations associated with both HI and free-free absorption are essentially featureless: they are both flat in the range $0 < \omega <  \dot{\theta}_v/\theta_{\rm max}$, and then either decline as $\omega^{\gamma-6}$ until $\omega >  \dot{\theta}_v/\theta_{\rm min}$ or they decline at $\omega > q_{\rm src} \dot{\theta}_v \approx \theta_{\rm src}^{-1} \dot{\theta}_v$, the point at which the finite source size hampers our ability to discern finer detail in the absorbing medium.  In practice, we certainly expect the background source to possess a greater angular size than $\theta_{\rm min}$, and possibly also $\theta_{\rm max}$: the $\omega^{\gamma-6}$ portion of the spectrum may not be discernible for all sources.

The ionized sheath model is distinguishable from the Gaussian model on temporal frequencies $\omega \ga \dot{\theta}_v/\theta_{\rm max}$, whereupon the power- and cross-spectra begin to oscillate on scales $\omega \sim \dot{\theta}_v / \theta_{\rm max} (1+\chi)$. For $\gamma < 3$ the amplitude of the envelope of oscillations in the power spectrum declines as $\omega^{-3}$ for $\omega> \theta_{\rm max}^{-1} \dot{\theta}_v$.

For sources whose angular sizes exceed the minimum cloud size ($\theta_{\rm src} > \theta_{\rm min}$), the expected minimum timescale of the variations associated with a medium that moves at an apparent speed $v_{\rm app}= D \dot{\theta}_v$ relative to a background source radiating at a brightness temperature $T_B$ is
\begin{eqnarray}
t_{\rm var} \sim \omega^{-1} = 9.9 \left( \frac{v_{\rm app}}{10^3\,{\rm km\,s}^{-1}} \right) \left( \frac{D}{100\,{\rm Mpc}} \right)
\left( \frac{S_\nu}{1\,{\rm mJy}}\right)^{1/2} \left( \frac{T_B}{10^{12}\,{\rm K}}\right)^{-1/2} \left(\frac{\nu}{1\,{\rm GHz}} \right)^{-1} \hbox{years},
\end{eqnarray}
and the corresponding minimum length scale probed is
\begin{eqnarray}
D \theta_{\rm src} = 1.0 \times 10^{-2}   \left( \frac{D}{100\,{\rm Mpc}} \right)
\left( \frac{S_\nu}{1\,{\rm mJy}}\right)^{1/2} \left( \frac{T_B}{10^{12}\,{\rm K}}\right)^{-1/2} \left(\frac{\nu}{1\,{\rm GHz}} \right)^{-1} \hbox{pc}.
\end{eqnarray}
Timescales less than a year are implied for some nearby bright sources, making objects such as Cen A, at a distance of $\approx 4\,$Mpc, particularly amenable to study.  Substantially faster variations are possible if the background source is the product of an outburst from an active galactic nucleus, and thus exhibits apparent superluminal motion, so that $v_{\rm app} \ga c$.  

If the linear extent of the background source is large compared to the outer scale of the power spectrum, equivalent to the largest cloud size in the discrete cloud models, the temporal spectrum of opacity-induced intensity fluctuations will be flat for {\it all} temporal frequencies up to a value $\omega \sim \dot{\theta}_v/\theta_{\rm src}$, and will cut-off sharply at higher frequencies.  This case applies to absorption against bright ($S_\nu>1\,$Jy) distant ($D>100\,$Mpc) radio sources with $v_{\rm app} \ll c$.  

Even though the detailed relationship between the spatial distribution of ionized and neutral hydrogen is only distinguishable for sources whose linear extents are smaller than the outer scale, it is always possible to at least determine the average relative neutral and ionization fractions of the absorbing medium.  The amplitudes of the power- and cross-spectra of temporal variations in the flat region of their spectra offer insight into the relationship between the two phases in an average sense.  Assuming identical values of $q_{\rm src}$ and ${\cal I}(\nu)$ at $\nu_{\rm HI}$ and $\nu_{\rm ff}$, the amplitudes of the temporal spectra in the limit $\omega \rightarrow 0$ reduce to:
\begin{subequations}
\begin{eqnarray}
{\cal P}(0,\nu_{\rm HI},\nu_{\rm HI})/{\cal P}(0,\nu_{\rm ff},\nu_{\rm ff}) &=&  
\frac{1}{n_0^2} \frac{\langle A_j^2 \rangle}{\langle B_j^2 \rangle} \begin{cases}
4, & \hbox{Gaussian model} \\
\chi^{-2} [3 + \chi (3 + \chi)]^{-2}, & \hbox{ionized sheath model} \\
\end{cases} \\
{\cal P}(0,\nu_{\rm HI},\nu_{\rm ff}) / {\cal P}(0,\nu_{\rm ff},\nu_{\rm ff}) &=& \frac{1}{n_0} \frac{\langle B_j^2 \rangle}{\langle A_j B_j \rangle} \begin{cases}
2^{-19/4}, & \hbox{Gaussian model} \\
\chi^{-1} [3 + \chi (3 + \chi)]^{-1}, & \hbox{ionized sheath model}. \\
\end{cases}
\end{eqnarray}
\end{subequations}
Numerically, the ratios involving the coefficients $A_j$ and $B_j$, which link the HI and free-free column densities of the $j$th cloud to its respective opacities, are given by,
\begin{subequations}
\begin{eqnarray}
\frac{1}{n_0^2} \frac{A_j^2}{B_j^2} &=& 
2.7 \times 10^{3}  \frac{X_j^2}{Y_j^2}  \left( \frac{n_0}{1\,{\rm cm}^{-3}} \right)^{-2} 
\left(\frac{\Delta v_{\rm eff}}{100\,{\rm km\,s}^{-1}} \right)^2 
\left( \frac{T_s}{10^3\,{\rm K}} \right)^{-2} 
 \left( \frac{T_e}{10^4\,{\rm K}} \right)^{2.7}
  \left( \frac{\nu_{\rm ff}}{1\,{\rm GHz}} \right)^{4.2} \qquad \hbox{and} \\
\frac{1}{n_0} \frac{B_j}{A_j} &=&  0.019 \frac{Y_j}{X_j}  \left( \frac{n_0}{1\,{\rm cm}^{-3}} \right)^{-1} 
\left(\frac{\Delta  v_{\rm eff}}{100\,{\rm km\,s}^{-1}} \right)^{-1}
\left( \frac{T_s}{10^3\,{\rm K}} \right) 
 \left( \frac{T_e}{10^4\,{\rm K}} \right)^{-1.35}
  \left( \frac{\nu_{\rm ff}}{1\,{\rm GHz}} \right)^{-2.1}, \
\end{eqnarray}
\end{subequations}
where $\langle X_j \rangle$ is the neutral fraction averaged all the clouds (subscripted by the index $j$) and $\langle Y_j \rangle$ is related to the ionization fraction of the clouds, as described in eqns.\,(\ref{Ajdefn}) and (\ref{Bjdefn}).  Thus we see that the relative amplitudes of the temporal power- and cross-spectra yield information on the ensemble-average neutral and ionized fractional content of the medium.

The spectra themselves offer information on the relative distribution of neutral and ionized matter only if the source size is smaller than the scale on which spatial differences in the distribution of absorbing material occurs; differences between the models are detectable only at high temporal frequencies. The Gaussian cloud model predicts a smooth power-law spectrum of temporal intensity fluctuations down to a scale set by the minimum cloud size, whereas the sharp ionized outer shells of the sheath model introduce features in the temporal power spectrum at scales $\omega \ga \dot{\theta}_v/[\theta_{\rm max} (1 + \chi)]$.  The most notable feature of the HI-free-free cross-power spectrum in the ionized sheath model is the sign change at a characteristic frequency given by the solution to the equation
\begin{eqnarray}
\omega^2 \approx \frac{3 }{4 \pi^2 \chi} \frac{\dot{\theta}_v^2}{\theta_{\rm max}^2} \left[ 2 + \cos \left( 4 \pi \omega \frac{\theta_{\rm max}}{\dot{\theta}_v} \right) \right],
\end{eqnarray}
which implies that the sign-flip occurs at  angular frequencies well in excess of $\dot{\theta}_v/\theta_{\rm max}$ (i.e. on timescales substantially shorter than the crossing time of the largest clouds).  

We stress that the underlying spatial power- and cross-spectra of HI and free-free opacity variations should be regarded as the fundamental quantities of interest.  In this sense the two cloud models considered in this paper represent extremes of a range of possible models for the spatial distribution of ionized and neutral material in the circumnuclear regions of external galaxies.  These models for the spatial distribution of matter provide a means of forward-modelling the temporal spectra, and their purpose is to provide a physical context in which to interpret the results of the temporal variations.

\subsection{Application to PKS\,1718-649}

As an illustration of the foregoing results, we consider the nearby ($z=0.0144$, $D=57.7\,$Mpc) GPS source PKS\,1718$-$649, whose changes in flux density and spectral shape in the range $1$-$3\,$GHz were interpreted as evidence of variations in the free-free opacity by \citet{tin15}.  The source was observed at three epochs over a 21 month period, and the 0.5\,Jy flux density change observed at 1 GHz over this period implies an opacity change of $\Delta \tau \ga 0.13$. Parkes observations at 725\,MHz and MWA observations at 199\,MHz further constrain the spectral shape and appear to exclude simple models that involve homogeneous absorbing material.  This provides strong motivation to consider whether spatial inhomogeneity in the free-free opacity in the foreground of the source is consistent with the temporal variations observed.

An estimate of the viability of this mechanism depends crucially on the actual timescale of the variations and on the source size, which determines the expected variability timescale.   The timescale of the variability is poorly constrained by only three observations, but it appears to be comparable to or to exceed the 10-month interval between the two closest observations.  The source size has not been measured in the range $1$-$3$\,GHz, but 4.8GHz VLBI imaging \citep{tin97} reveals a double-lobed source with the NW and SE lobes having dimensions of $2.3 \times 1.2\,$mas and $2.4 \times 1.5\,$mas and brightness temperatures of $5 \times 10^{10}\,$K and $2 \times 10^{10}\,$K respectively; 22\,GHz imaging reveals a similar structure with comparable lobe sizes \citep{tin03}.  At the distance of the source 1\,mas corresponds to a linear scale of 0.28\,pc.  We place a lower limit on the the expected crossing time of
\begin{eqnarray}
1.1 \left( \frac{\theta_{\rm src}}{1.2\,{\rm mas}} \right) \left( \frac{v_{\rm app}}{c} \right)^{-1} \,\hbox{years,}
\end{eqnarray}
where we have normalised the source size to the angular size of the minor axis of the NW lobe and where $v_{\rm app}$ is the apparent relative speed between the emitting material and the foreground absorbing medium. The low brightness temperatures of the components of the sources relative to the Inverse Compton limit suggests that the emission is not relativistically beamed.  Thus the apparent crossing velocity of material across the source is likely $v_{\rm app} < c$.  A timescale much greater than a year is expected if the jet is highly frustrated by dense material along its axis, so that the bulk of the emission moves slowly across the emitting source. 
An upper limit to the timescale is obtained by assuming an apparent relative speed $v_{\rm app}  \sim 10^{3}\,$km\,s$^{-1}$, characteristic for absorbing clouds orbiting a stationary radio-emitting nuclear region, yielding an expected variability timescale of $3\times 10^2$\,years.  Clearly, multi-epoch VLBI measurements of the source structure are necessary to place limits on the apparent  speed of emitting material within the source.  Thus, although present data is insufficient to predict a variability timescale, we can at least conclude that the model is consistent with present constraints on the observed variations.   




%



\section{Conclusion} \label{sec:Conclusions}
We have presented a formalism for interpreting the temporal variations in HI and free-free opacity observed in some young, compact radio sources and for connecting these with the underlying properties of the circumnuclear media of their host galaxies.
The theory is, however, more widely applicable to any medium comprised of ionized and neutral material drifting in front of a radio source at a fixed velocity. 

The background source structure acts as a low pass filter on the opacity variations observable against these sources and limits the timescales of variations and minimum size of the structures probed within the medium.  For drift rates of $\sim 10^3\,$km\,s$^{-1}$ against sources of brightness temperatures $\sim 10^{12}\,$K the expected timescale of variability is of order years to decades.  Much faster variations are possible against radio-emitting outbursts with apparently superluminal speeds, giving rise to variations on timescales $\ga 300$ times smaller (i.e. of order days to weeks). However, such motions are not thought to be relevant to some young radio sources in which free-free opacity variations are believed to occur \citep[e.g.,][]{tin15}.  The expected timescale is potentially an important distinguishing feature of this model relative to other proposed variability mechanisms, particularly those related to interstellar scintillation, which typically gives rise to variations on day to several-week timescales; apparent absorption variations can result from the scintillation of the flux density associated with any fine-scale angular structure imprinted in the image of a source by structure in the foreground absorbing medium. We refer the reader to \citet{mac05} for a more detailed discussion as it applies to several-week timescale variations in HI.

Measurements of the power- and cross-power spectrum of intensity fluctuations associated with the HI and free-free opacity variations can resolve the nature of the spatial relationship between the two phases.  We have investigated two contrasting models for their distribution --- one based on clouds with Gaussian density profiles, and another based on clouds comprised of neutral cores with ionized sheaths --- and shown how the power spectra of temporal intensity fluctuations differs between the two cases.  However, source structure limits the angular scales accessible to $\ga \theta_{\rm src}$ so that, if the ionized and neutral medium is predominately distinguishable only on scales smaller than this, then these scales will not be accessible through measurements of temporal opacity fluctuations.  However, even in this case it is still possible to discern details about the average ionization and neutral fraction of the medium through comparison of the HI and free-free opacity temporal variations.  Specifically, comparison of the amplitudes of the power spectra of intensity fluctuations caused by these two absorption mechanisms reveals the ratio $n_0^{-2} \langle X^2 \rangle/\langle Y^2 \rangle$, where $X$ is the neutral fraction and $Y$ is the ionization fraction relative to the overall density, $n_0$.  The amplitude of the cross-power spectrum between HI and free-free intensity variations relative to the free-free power spectrum further reveals the quantity $n_0^{-1} \langle Y \rangle/\langle X \rangle$.

Finally, we note that there are exciting prospects for undertaking future investigations of the interstellar media of young compact extragalactic radio sources.  Facilities such as the MWA \citep{tin13,lon09} are capable of making extremely sensitive measurements of variations in free-free opacity at the low frequencies, $\sim 80-250\,$MHz, at which the effects of free-free absorption are easily measurable.  Meanwhile, cm-wavelength facilities such as ASKAP \citep{jon08,jon07} are capable of undertaking large surveys for radio galaxies with variable HI absorption \citep{all15} and, with its high spectral stability and dynamic range, in measuring extremely small opacity variations.

\section*{Acknowledgments}
Parts of this research were conducted by the Australian Research Council Centre of Excellence for All-sky Astrophysics (CAASTRO), through project number CE110001020.

\appendix

\section{The power and cross-power spectra of clouds with a neutral core and ionized sheath} \label{app:Sheath}
Here we present the analytic forms of the the power spectra of the optical depth variations for clouds comprised of a constant-density neutral core and a constant-density ionized sheath for the model presented in \S \ref{subsec:IonizedSheath}.  For a power law distribution of cloud sizes with index $-\gamma$ and for cloud sizes ranging from $\theta_{\rm min}$ to $\theta_{\rm max}$ we obtain the following expressions:
\begin{eqnarray}
\Phi_\tau ({\bf q};\nu_{\rm HI},\nu_{\rm HI}) &=& \frac{D^2 n_0^2 N \langle A_j^2 \rangle}{4 \pi^4 q^6} \int_{\theta_{\rm min}}^{\theta_{\rm max}} d\theta_c \, p(\theta_c) \theta_c^2 \left[ \frac{\sin (2 \pi q \theta_c)}{\theta_c} - 2\pi q \cos (2 \pi q \theta_c) \right]^2 \nonumber \\
&=& \frac{D^2 n_0^2 N K \langle A_j^2 \rangle }{64 \pi^4 q^6 }
{\Bigg \vert} 
\theta_c^{1-\gamma} {\Bigg [} \frac{8 (3 -\gamma - 4\pi^2 q^2 (\gamma-1) \theta_c^2)}{(\gamma-1)(\gamma-3)} - 2 (\gamma-6) \cos(4 \pi q \theta_c) + 8 \pi q \theta_c \sin (4 \pi q \theta_c)  \nonumber \\
&\null&  \quad + (\gamma-5)(\gamma-2)  [{\rm E}_\gamma (-4 i \pi q \theta_c)+  {\rm E}_\gamma (4 i \pi q \theta_c)] 
{\Bigg ]} 
{\Bigg \vert}_{\theta_c=\theta_{\rm min}}^{\theta_c=\theta_{\rm max}},
\end{eqnarray}
\begin{eqnarray}
\Phi_\tau ({\bf q};\nu_{\rm ff},\nu_{\rm ff}) &=& N  \langle B_j^2  \rangle \int_{\theta_{\rm min}}^{\theta_{\rm max}} d\theta_c \, p(\theta_c) \theta_c \tilde{{\cal T}}_{\rm ff}^2 ({\bf q};\theta_c) \nonumber \\
&=& \frac{D^2 n_0^4  N K \langle B_j^2  \rangle}{64 \pi^4 q^6 \, \chi^2 (2+\chi)^2 (\gamma-1)(\gamma-2)(\gamma-3)}  \, {\Bigg \vert} \theta^{1-\gamma} {\Bigg [}
e^{4 \pi i q \theta} (6- \gamma-4 i \pi q \theta) + (\gamma-5)(\gamma-2) [E_\gamma (4 \pi i q \theta) + E_\gamma (-4\pi i q \theta)] 
\nonumber \\
&\null& - \frac{8 (\gamma-2) ( (\gamma-1)(1+ \chi) + \chi^2)}{\chi^2} [E_\gamma (2 \pi i q \theta \chi) + E_\gamma (-2 \pi i q \theta \chi) ] 
\nonumber \\ &\null&
+ (\gamma-5) (\gamma-2) [E_\gamma (4 \pi i q \theta (1+\chi)) + E_\gamma (-4 \pi i q \theta (1+\chi))] \nonumber \\
&\null& 
- \frac{8 (\gamma-2) ((1+\chi)(\gamma-5) - \chi^2)}{(2+\chi)^2} [E_\gamma(2 \pi i q \theta(2+\chi)) + E_\gamma (-2 \pi i q \theta(2+\chi))]
\nonumber \\ &\null&
+ \left[\frac{i (\gamma-8) \gamma}{\gamma-2} + 4\pi q \theta \right] \sin(4 \pi q \theta) 
+ \frac{8 \pi q \theta (1+\chi)}{\chi} \left[\chi \sin(4 \pi q \theta (1+\chi)) - 4 \sin (2 \pi q \theta \chi) \right]  \nonumber \\ &\null&
-16 \chi ^2 (\gamma -2)  \left\{ (\chi +2)^2 \left[ \gamma  \left(2 \pi ^2 \theta ^2
   q^2 (\chi  (\chi +2)+2)+1\right)-2 \pi ^2 \theta ^2 q^2 (\chi  (\chi
   +2)+2)-3 \right]  \right. \nonumber \\ &\null& \left. 
   +2 \pi  \theta  q (\gamma -3) (\gamma -1)  (\chi +1) (\chi +2) \sin (2 \pi  \theta  q (\chi +2)) 
   + (\gamma -3) (\gamma -1) ((6-\gamma)(1+\chi) + \chi^2) \cos (2 \pi  \theta  q (\chi +2)) \right\}  
   \nonumber \\ &\null&
    + (\gamma -3) (\gamma -1) (\chi +2)^2 \left\{ (\gamma -2) \left[ \chi ^2 (6-\gamma +4 i \pi  \theta  q) \cos (4 \pi  \theta  q)
    -2 (\gamma -6) \chi ^2 \cos (4 \pi \theta  q (\chi +1)) \right. \right. \nonumber \\ &\null& \left. \left.
    +16 \left((\gamma -2) \chi +\gamma +\chi ^2-2\right) \cos
   (2 \pi  \theta  q \chi )\right]+12 i \chi ^2 \sin (4 \pi  \theta  q)\right\}
{\Bigg ]}
{\Bigg \vert}_{\theta_c=\theta_{\rm min}}^{\theta_c=\theta_{\rm max}},
\end{eqnarray}
and
\begin{eqnarray}
\Phi_\tau ({\bf q};\nu_{\rm HI},\nu_{\rm ff}) &=&  N \langle A_j B_j \rangle \int_{\theta_{\rm min}}^{\theta_{\rm max}} d\theta_c \, p(\theta_c) \theta_c \tilde{{\cal T}}_{\rm HI} ({\bf q};\theta_c) \tilde{{\cal T}}_{\rm ff} ({\bf q};\theta_c) \nonumber \\
&=& \frac{D^2 n_0^3 N K \langle A_j B_j \rangle}{16 \pi^4 q^6} {\Bigg \vert} \theta^{1-\gamma} {\Bigg [}
\frac{2}{\gamma-1} + \frac{8 \pi^2 q^2 \theta^2}{\gamma-3} - \frac{\gamma (1+\chi) -6 - \chi (\chi+6) - 2 \pi i q \theta (1+\chi)(2+\chi) }{(2+\chi)^2} e^{-2 \pi i q \theta (2+\chi)} \nonumber \\ &\null&
- \frac{\gamma (1+\chi) - 6 - \chi (\chi+6) + 2 \pi i q \theta (1+\chi)(2+\chi)}{(2+\chi)^2} e^{2 \pi i q \theta (2+\chi)} + 4 \pi^2 q^2 \theta^2 E_{\gamma-2} (-4\pi i q \theta) \nonumber \\ &\null&
+ 4 \pi^2 q^2 \theta^2 \left( \frac{\gamma-6}{\gamma-2} \right) E_{\gamma-2}(4 \pi i q \theta) 
- 4 \pi^2 q^2 \theta^2 \left(1 + \chi + \frac{\chi^2}{\gamma-2} \right) [E_{\gamma-2}(2 \pi i q \theta \chi) + E_{\gamma-2} (-2 \pi i q \theta \chi)] \nonumber \\ &\null&
+ 4 \pi i q \theta E_{\gamma-1} (-4 \pi i q \theta) - [E_\gamma (4 \pi i q \theta) + E_\gamma (-4 \pi i q \theta)] 
- [E_\gamma (2 \pi i q \theta \chi) + E_\gamma (-2 \pi i q \theta \chi) ] \nonumber \\ &\null&
- \frac{(\gamma-2) ( (5-\gamma)(1+\chi) + \chi^2)}{(2+\chi)^2} [E_\gamma (2 \pi i q \theta(2+\chi)) + E_\gamma (2 \pi i q \theta (2+\chi))]
\nonumber \\ &\null& 
- \frac{4 \pi q \theta}{\gamma-2} \left[ \sin(4 \pi q \theta) + \chi \sin (2 \pi q \theta \chi) + i \cos (4 \pi q \theta) \right] 
{\Bigg ]}
{\Bigg \vert}_{\theta_c=\theta_{\rm min}}^{\theta_c=\theta_{\rm max}}.
\end{eqnarray}
The asymptotic behaviour of these results may be derived by noting that $E_\gamma(x) \rightarrow e^{-z}/z$ as $z \rightarrow \infty$.  Terms of the form $E_{\gamma} (-4 i z) + E_\gamma (4 i z)$ reduce to $-2 \sin(z)/z$ in the limit $z \rightarrow \infty$.   For example, the power spectrum of HI opacity fluctuations reduces to  
\begin{eqnarray}
\Phi_\tau (q,\nu_{\rm HI},\nu_{\rm HI}) \rightarrow \frac{D^2 n_0^2 N K \langle A_j^2 \rangle}{8 \pi^4 q^6} \left[ \frac{\theta_c^{1-\gamma}}{1-\gamma} + \frac{4 \pi^2 q^2 \theta_c^{3-\gamma}}{3-\gamma} \right]_{\theta_{\rm min}}^{\theta_{\rm max}}
\end{eqnarray}
in this limit.  We obtain the behaviour of the spectra in the limit $q \rightarrow 0$ by expanding their numerators to ${\cal O}(q^7)$:
\begin{subequations}
\begin{eqnarray}
\lim_{q \rightarrow 0} \Phi_\tau ({\bf q};\nu_{\rm HI},\nu_{\rm HI}) &=& D^2 n_0^2 K \langle A_j^2 \rangle \frac{16 \pi^2 \theta_{\rm max}^{7-\gamma}}{9 (7 - \gamma)},   \\ 
\lim_{q \rightarrow 0} \Phi_\tau ({\bf q};\nu_{\rm ff},\nu_{\rm ff}) &=& D^2 n_0^4 K \langle B_j^2 \rangle \frac{16 \pi^2 \theta_{\rm max}^{7-\gamma} }{9 (7-\gamma)} \chi^2 [3 + \chi (3+\chi)]^2,   \\
\lim_{q \rightarrow 0} \Phi_\tau ({\bf q};\nu_{\rm HI},\nu_{\rm ff}) &=& D^2 n_0^3 K \langle A_j B_j \rangle \frac{16 \pi^2 \theta_{\rm max}^{7-\gamma} }{9 (7-\gamma)} \chi [3 + \chi (3+\chi)].
\end{eqnarray}
\end{subequations}

\section{The temporal power spectrum} \label{app:PowSpec}
In this Appendix we compute the temporal power spectrum associated with a medium whose spatial opacity power spectrum follows a power law, of the form given in eq.(\ref{GenericPowerLaw}).  In this treatment we consider the effect of finite source size by modelling the source as a gaussian which cuts the spatial power spectrum off at a scale $q_{\rm max}$, as discussed in \S \ref{sec:TempSpec} above.

The evaluation of the temporal power spectrum is broken into two regions, one for  $\omega/\dot{\theta}_v > q_{\rm min}$, and the other for low angular frequencies $\omega/\dot{\theta}_v < q_{\rm min}$:
\begin{eqnarray}
\langle {\cal P} \rangle (\omega)  &=& \frac{2 Q_0 {\cal I}(\nu) {\cal I}(\nu')}{ \dot{\theta}_v} \int_{0}^{\infty}  \left[ q_\perp^2 + \left( \frac{\omega}{\dot{\theta}_v} \right)^2 \right]^{-\beta/2}  \exp \left[- \frac{(\omega/\dot{\theta}_v)^2+ q_\perp^2}{q_{\rm src}^2} \right] dq_\perp,  \quad \omega/\dot{\theta}_v > q_{\rm min} \nonumber \\
&=& \frac{Q_0 {\cal I}(\nu) {\cal I}(\nu') \pi}{ \dot{\theta}_v} \sec \left( \frac{\pi \beta}{2} \right) \left[ q_{\rm src}^{1-\beta} 
\null_1 \tilde{F}_1 \left( \frac{1}{2};\frac{\beta+1}{2}; - \frac{(\omega/\dot{\theta}_v)^2}{q_{\rm src}^2} \right) 
- \frac{\sqrt{\pi}}{\Gamma(\beta/2)} \left( \frac{\omega}{\dot{\theta}_v} \right)^{1-\beta} 
\null_1 \tilde{F}_1 \left( \frac{2-\beta}{2};\frac{3-\beta}{2}; - \frac{(\omega/\dot{\theta}_v)^2}{q_{\rm src}^2} \right) \right], \nonumber \\
&\null& \hskip 13cm  \omega/\dot{\theta}_v > q_{\rm min}.
\end{eqnarray}
where $\null_1 \tilde F_1(a,b;z)$ is the regularized confluent hypergeometric function $\null_1 F_1(a;b;z)/\Gamma(b)$.
The low angular frequency component of the spectrum is 
\begin{eqnarray}
\langle {\cal P} \rangle (\omega) &=& 
\frac{2 {\cal I}(\nu) {\cal I}(\nu') Q_0}{ \dot{\theta}_v} \left\{ 
\int_0^{\sqrt{q_{\rm min}^2 - (\omega/\dot{\theta}_v)^2} } 
 q_{\rm min}^{-\beta} \exp \left[- \frac{(\omega/\dot{\theta}_v)^2+ q_\perp^2}{q_{\rm src}^2} \right] dq_\perp \right. \nonumber \\
&\null& \left. \qquad +  \int_{\sqrt{q_{\rm min}^2 - (\omega/\dot{\theta}_v)^2}}^{\infty} \left[ q_\perp^2 + \left( \frac{\omega}{\dot{\theta}_v} \right)^2 \right]^{-\beta/2}  \exp \left[- \frac{(\omega/\dot{\theta}_v)^2+ q_\perp^2}{q_{\rm src}^2} \right] dq_\perp \right\}, \qquad \omega/\dot{\theta_v} < q_{\rm min}.
\end{eqnarray}
The second integral is intractable, so we approximate it by replacing the exponential source-related term with an upper cutoff at $q_\perp = \sqrt{q_{\rm src}^2 - (\omega/\dot{\theta}_v)^2}$ to obtain
\begin{eqnarray}
\langle {\cal P} \rangle (\omega) &\approx& 
\frac{{\cal I}(\nu) {\cal I}(\nu') Q_0}{ \dot{\theta}_v}  \left\{ \pi^{1/2} q_{\rm min}^{-\beta} q_{\rm src} {\rm erf} \left( \frac{\sqrt{q_{\rm min}^2 - (\omega/\dot{\theta}_v)^2}}{q_{\rm src}} \right) \exp \left( - \frac{(\omega/\dot{\theta}_v)^2}{q_{\rm src}^2} \right) \right. \nonumber \\
&\null& \left. \quad
+ \frac{2}{\beta-1} H (q_{\rm src} - q_{\rm min}) \left[ 
[q_{\rm min}^2 - (\omega/\dot{\theta}_v)^2]^{(1-\beta)/2} \null_2 F_1 \left(\frac{\beta-1}{2},\frac{\beta}{2} ; \frac{1+\beta}{2}; \frac{(\omega/\dot{\theta}_v)^2}{(\omega/\dot{\theta}_v)^2 - q_{\rm min}^2} \right) \right. \right. \nonumber \\ 
&\null& \left. \left. \qquad \qquad -   [q_{\rm src}^2 - (\omega/\dot{\theta}_v)^2]^{(1-\beta)/2} \null_2 F_1 \left(\frac{\beta-1}{2},\frac{\beta}{2} ; \frac{1+\beta}{2}; \frac{(\omega/\dot{\theta}_v)^2}{(\omega/\dot{\theta}_v)^2 - q_{\rm src}^2} \right)
\right]   
\right\}, \qquad  \omega/\dot{\theta}_v < q_{\rm min}.
\end{eqnarray}
In the limit $\omega \rightarrow 0$ the contributions proportional to  the two hypergeometric functions cancel and we obtain
\begin{eqnarray}
\lim_{\omega \rightarrow 0} \langle P(\omega) \rangle = \frac{{\cal I}(\nu) {\cal I}(\nu') Q_0}{ \dot{\theta}_v} 
 \pi^{1/2} q_{\rm min}^{-\beta} q_{\rm src} {\rm erf} \left( \frac{q_{\rm min}}{q_{\rm src}} \right).
\end{eqnarray}

\end{document}